\documentclass[10pt]{article}
\usepackage{amsmath,amsfonts,vmargin}
\usepackage{amsbsy}
\usepackage{verbatim}
\usepackage{setspace}
\usepackage{rotating}
\usepackage{array}
\usepackage{pdflscape}

\newcommand{\bz}{{\bf z}}
\newcommand{\bzro}{{\bf 0}}

\newcommand{\bSig}{\mbox{\boldmath{$\Sigma$}}}
\newcommand{\bgam}{\mbox{\boldmath{$\gamma$}}}

\newcommand{\bomg}{\mbox{\boldmath{$\Omega$}}}
\newcommand{\bLam}{\mbox{\boldmath{$\Lambda$}}}
\newcommand{\bQ}{\mbox{\boldmath{$Q$}}}

\newcommand{\bxi}{\mbox{\boldmath{$\xi$}}}
\newcommand{\bvarphi}{\mbox{\boldmath{$\varphi$}}}

\newcommand{\btU}{\mbox{\boldmath{$\tilde{U}$}}}

\newcommand{\bZ}{{\bf Z}}

\newcommand{\bV}{{\bf V}}

\newcommand{\bC}{{\bf C}}

\newcommand{\bU}{{\bf U}}
\newcommand{\bR}{{\bf R}}
\newcommand{\bJ}{{\bf J}}

\newcommand{\Var}{\mbox{Var}}

\newcommand{\mc}{\multicolumn}
\newcommand{\real}{\mathbb{R}}

\newcommand{\bth}{{\mbox{\boldmath{$\theta$}}}}
\newcommand{\bthsub}{{\mbox{\scriptsize \boldmath{$\theta$}}}}
\newcommand{\bThsub}{{\mbox{\scriptsize \boldmath{$\Theta$}}}}
\newcommand{\tsub}{{\mbox{\scriptsize {$t$}}}}
\newcommand{\dsub}{{\mbox{\scriptsize {$d$}}}}
\newcommand{\alphsub}{{\mbox{\scriptsize {$\alpha$}}}}
\newcommand{\bTh}{{\mbox{\boldmath{$\Theta$}}}}

\newcommand{\bps}{{\mbox{\boldmath{$\psi$}}}}
\newcommand{\bpsi}{{\mbox{\boldmath{$\psi$}}}}
\newcommand{\bPs}{{\mbox{\boldmath{$\Psi$}}}}

\newcommand{\bhpsi}{\mbox{\boldmath{$\hat{\psi}$}}}

\newcommand{\cA}{\mathcal{A}}
\newcommand{\cB}{\mathcal{B}}
\title{Cox Model with Covariate Measurement Error and Unknown Changepoint}
\author{Sarit Agami\\
Department of Statistics\\
Hebrew University, Mount Scopus, Jerusalem, Israel\\
email:sarit.agami@mail.huji.ac.il\\
\and
David M. Zucker\\
Department of Statistics\\
Hebrew University, Mount Scopus, Jerusalem, Israel\\
\and
Donna Spiegelman\\
Departments of Epidemiology, Biostatistics, Nutrition and Global Health\\
Harvard T.H. Chan School of Public Health, Boston MA, USA}

\date{\today}

\begin{document}
\maketitle




\begin{abstract}
The standard Cox model in survival analysis assumes that the covariate effect is constant across the entire covariate domain.
However, in many applications, there is interest in considering the possibility
that the covariate of main interest is subject to a threshold effect: a change in the slope at a certain
point within the covariate domain. Often, the value of this threshold is unknown and need to be estimated. In addition, often, the covariate of interest is not measured exactly, but rather is subject to some degree of measurement error. In this paper, we discuss estimation of the model parameters under an independent additive error model where the covariate of interesting is measured with error and the potential threshold value in this covariate is unknown. As in earlier work which discussed the case of konwn threshold, we study the performance of several bias correction methods: two versions of regression calibration (RC1 and RC2), two versions of the fitting a model for the induced relative risk (RR1 and RR2), maximum pseudo-partial likelihood estimator (MPPLE) and simulation-extrapolation (SIMEX). These correction methods are compared with the naive estimator. We develop the relevant theory, present a simulation study comparing the several correction methods, and illustrate the use of the bias correction methods in data from the Nurses Health Study (NHS) concerning the relationship between chronic air pollution exposure to particulate matter of diameter 10 $\mu$m or less (PM$_{10}$). The simulation results suggest that the best overall choice of bias correction method is either the RR2 method or the MPPLE method.
\end{abstract}

\maketitle

\renewcommand{\baselinestretch}{1.5}

\newpage
\normalsize

\section{Introduction \label{section intro}}
The Cox regression model with a threshold effect and a measurement error in the main covariate, was described by Agami et al. (2018).
Let $X(t)$ denote the covariate of main interest and $\bZ(t) \in \real^p$ the vector of additional covariates.
The main covariate $X(t)$ is subject to measurement error, while the additional covariates $\bZ(t)$ are error-free.
The measurement error in $X(t)$ is assumed to be additive. That is, the observed covariate is $W(t) = X(t) + U$, where $U$ is a random
variable such that $E(U|X(t))=0$. Define $u_+ = \max(u,0)$. The model is then given by
\begin{equation}
\lambda(t|x(t),\bz(t)) = \lambda_0(t) \exp(\bgam^T \bz(t) + \beta x(t) + \omega (x(t)-\tau)_+).
\label{CP1}
\end{equation}
Agami et al. (2018) considered the case where the changepoint is known. Often, this changapoint is unknown, and in this paper we consider this case. We examine the methods for measurement error correction which were described and examined by Agami et al. (2018).
We seek to estimate $\beta$, $\omega$, $\bgam$ and $\tau$.
Section 2 presents the notation and background, and give a short description of the methods examined. Section 3 presents the asymptotic properties of the methods.
Section 4 presents a simulation study comparing the various methods. Section 5 presents real example using data from the Nurses' Health Study
(NHS) on the relationship between air pollution, expressed in terms of exposure to particulate matter of diameter 10 $ug$/$m^3$ or less (PM$_{10}$),
and fatal myocardial infarction. Section 6 presents a brief summary and discussion, and Section 7 presents the technical proofs.

\section{Methods Considered}

\subsection{Setting, Notation, and Background}
We consider the standard survival analysis setup with right censoring, where we also allow for left-truncation.
The observations are on $n$ independent individuals. For a given individual $i$, $(\bZ_i(t),X_i(t))$ denotes the true
covariate vector, $\tilde{T}_i$ denotes the time of entry into the study, $T_i^\circ$ denotes the survival time,
and $C_i$ denotes the time of right censoring. We work with the classical normal additive measurement error model, that is,
$W_i(t) = X_i(t) + U_i$, where the conditional distribution of $X_i(t)$ given
$\bZ_i(t)=z$ is $N(\mu_x(z),\sigma_x^2)$ and the $U_i$'s are i.i.d.\
$N(0,\sigma_u^2)$, independent of the $X(t)$'s and the $\bZ(t)$'s.
We assume that $\mu_x(z)$ is of the form $\mu_x(z) = \alpha_0 + \alpha_1 z$.
We further assume that $(\tilde{T}_i, C_i)$
is conditionally independent of $(X_i(t), T_i^\circ)$ given $\bZ_i(t)$.
The observed data consist of $(W_i(t),\bZ_i(t))$, the entry time $\tilde{T}_i$, the observed follow-up time $T_i=\min(T_i^\circ,C_i)$, and the event indicator $\delta_i = I(T_i^\circ \leq C_i)$, where the survival time $T_i^\circ$ follows the model (\ref{CP1}).
Let denote the event counting process by $N_{i} \left(t\right)=I\left(T_{i}^{} \le t,\delta _{i} =1\right)$,
and the at-risk indicator by $Y_i(t)=I(\tilde{T}_i \leq t \leq T_i)$. Also denote
$\lambda _{i} \left(t\right)dt=\lambda _{0} \left(t\right)\exp \left(\beta X_{i} \left(t\right)+\omega \left(X_{i} \left(t\right)-\tau \right)_{+} +\bgam ^{T} \bZ_{i} \left(t\right)\right)dt$, $\tilde{\lambda }_{i} \left(t\right)dt=Y_{i} \left(t\right)\lambda _{i} \left(t\right)dt$ and $\bar{\tilde{\lambda }}\left(t\right)=\sum _{i=1}^{n}\tilde{\lambda }_{i} \left(t\right)$.
We define $d\tilde{F}\left(t\right)=E\left[Y_{i} \left(t\right)\lambda _{i} \left(t\right)\right]dt$.
The maximum possible follow-up time is denoted by $t^{*}$.
We write $\bth=(\bps ,\tau)$ where $\bps = \left(\bgam ^{T} ,\beta ,\omega \right)$, and we denote $\sigma_w^2 = \Var(W(t)|\bZ(t)) = \sigma_x^2 + \sigma_u^2$.
As in Agami et al. (2018), we consider in our simulation work the simple setting without additional covariates
$\bZ(t)$, and assume that $\mu_x=\alpha_0, \sigma_x^2$, and $\sigma_u^2$ are estimated based on an external replicate measures study. The estimates are computed by one-way random effects ANOVA, and are assumed transportable to the main study.

\subsection{Description of the Methods Examined}
Let us write the relative risk as $r(x,z,\bth) = \exp(\bgam^T \bz + \beta x + \omega (x - \tau)_+)$. If $X(t)$ were known, the standard Cox log partial likelihood is given by
\begin{align*}
l_{p} \left(\bth \right) & = \sum _{i=1}^{n} \delta _{i}  \left[ \log r(X_i(t),\bZ_i(t),\bth)
-\log \sum _{j=1}^{n}Y_{j}(T_{i}) r(X_i(t),\bZ_i(t),\bth) \right] \\
& = \sum _{i=1}^{n} \delta _{i} \left[ (\bgam^T \bZ_i(t) + \beta X_i(t) + \omega (X_i(t) - \tau)_+)
- \log \sum _{j=1}^{n}Y_{j}(T_{i}) r(X_i(t),\bZ_i(t),\bth) \right].
\end{align*}

\noindent We assume the following assumptions throughout the paper.

\noindent GA1. The parameter space $\bTh $ is compact, and $\bth ^{*} $, which will be defined for each method, is an interior point.

\noindent GA2. The vector $W\left(t\right)$ is left-continuous,
the vector of the additional risk factors $\bZ\left(t\right)$ is bounded and left-continuous,
and $P\left(Y_{i} \left(t^{*} \right)>0\right)>0$.

\noindent GA3. $\lambda _{0} \left(t\right)$ is bounded.

\noindent GA4. There exists a set of fixed times $\kappa _{1} ,...,\kappa _{R}$ and a set of random variables
$Q_{ijr} ,\, \, i=1,\, ...,\, n,\, \, j=1,\, ...,\, p,\, \, r=1,\, ...,\, R$ and a continuous function $G$ such
that $X_{i} \left(t\right)=G\left(t,\, Q_{ir}^{*} \left(t\right)\right)$, where $Q_{ir}^{*} =
\left\{Q_{ijs} \, :\, j=1,\, ...,\, p,\, s=1,\, ...,\, r(t)\right\}$ and $r\left(t\right)=
\max \left\{r:\kappa _{r} \le t\right\}$.

Conditions GA1-GA3 are standard. Condition GA4 is needed to apply the functional central limit theorem to
certain quantities involved in the objective function used in the estimation.

Generally, the methods for Cox regression analysis with covariate error involve replacing $r(x,z,\bth)$
with some substitute. The specific methods we examine are listed below. This is a short description only, for more
details see Agami et al.(2018).


\noindent\textbf{A. Naive Method}: $X_{i}(t)$ is replaced with $W_{i}(t)$.

\noindent \textbf{B. Regression Calibration (RC) Methods}

\noindent \textbf{B1.  Simple RC Method (RC1)}: $X_{i}(t)$ is replaced with
$\mu(W_i(t),\bZ_i(t))=E\left(\left.X_{i}(t) \right|W_{i}(t), \bZ_i(t) \right)$.
\noindent \textbf{B2.  Improved RC Method (RC2)}: $X_{i}(t)$ is replaced with $E\left(\left. X_{i}(t) \right|W_{i}(t),
\bZ_i(t) \right)$ and $\left(X_{i}(t) -\tau \right)_{+} $ is replaced with $E\left(\left. \left(X_{i}(t) -\tau \right)_{+}
\right|W_{i}(t), \bZ_i(t) \right)$.
\\ \textbf{C. Induced Relative Risk (RR) Methods}

\noindent \textbf{C1. Original RR Method (RR1)}: $\exp (\beta X(t)+\omega (X(t)-\tau )_{+})$ is replaced with
$E[\exp (\beta X(t)+\omega (X(t)-\tau )_{+}) |W(t)=w, \bZ(t)=z]$.

\noindent \textbf{C2. Modified RR Method (RR2)}: This is a version of RR1 which involves
a bootstrap bias-correction procedure.

\noindent \textbf{D. MPPLE Method}: The MPPLE method of Zucker (2005) involves substituting the induced hazard into the
Cox partial likelihood and maximizing over $\bth$. The induced hazard depends on the unknown cumulative hazard rate $\Lambda_0(t)$.

\noindent \textbf{E. SIMEX Method}: In preliminary work, we examined three extrapolation methods: rational linear extrapolation, simple quadratic extrapolation, and the third-degree polynomial extrapolant. Examining plots with the fitted
extrapolation function superimposed on a scatterplot of the mean value of the estimate (based on 1000 replications),
we found that the first-degree polynomial provided the best fit, and we used this extrapolation method in the implementation of the SIMEX estimator in our numerical studies.
\section{Asymptotic properties of the Naive, RC1, RC2, RR1, and MPPLE estimators}
\subsection{Description}
 For $\tau $ unknown, with the naive and RC1 methods, the log partial likelihood is not differentiable respect to $\tau $, but the limit of the log partial likelihood is differentiable in $\tau $, as we show below. For the RC2, RR and MPPLE methods, the log-likelihood is differentiable in $\tau $, and so classical asymptotic theory goes through in a standard way.
\subsection{The Naive and RC1 Methods}
\noindent The naive and the RC1 estimators are of a common form. Each involves replacing of $X_{i} \left(t\right)$ in the term
$\beta X_{i} \left(t\right)$ by a surrogate $g_{1} \left(W_{i} \left(t\right),\, \bZ_{i} \left(t\right),\tau \right)$ and
$\left(X_{i} \left(t\right)-\tau \right)_{+} $ in the term $\omega \left(X_{i} \left(t\right)-\tau \right)_{+} $ by a surrogate
$g_{2} \left(W_{i} \left(t\right),\, \bZ_{i} \left(t\right),\tau \right)$: the naive method takes $g_{1} \left(w,\, z,\tau
\right)=w$ and $g_{2} \left(w,\, z,\tau \right)=\left(w-\tau \right)_{+} $, and the RC1 method takes $g_{1} \left(w,\, z,\tau
\right)=\mu \left(w,z\right)$ and $g_{2} \left(w,z\right)=\left(\mu \left(w,z\right)-\tau \right)_{+} $. Let $g$ denote the
function pair $\left(g_{1} ,g_{2} \right)$ and let $\bV_{i} \left(g\left(t,\tau \right)\right)$ denote a vector of length $p+2$ in
which the first $p$ components are the elements of $\bZ_{i} \left(t\right)$, the $\left(p+1\right)-$th component is $g_{1}
\left(W_{i} \left(t\right)\, ,\bZ_{i} \left(t\right),\tau \right)$, and the $\left(p+2\right)-$th component is $g_{2} \left(W_{i}
\left(t\right)\, ,\, \bZ_{i} \left(t\right),\tau \right)$.

\noindent Then the log partial likelihood is
\begin{align*}
l_{p} \left(\bth ,g\right) = \sum _{i=1}^{n} \delta _{i} \left[ \bps ^{T} \bV_{i} \left(g\left(t,\tau \right)\right)\right)]
- \log \sum _{j=1}^{n}Y_{j}(T_{i})  \left[ \bps ^{T} \bV_{i} \left(g\left(t,\tau \right)\right)\right)].
\end{align*}
Denote $\bxi _{i} \left(t,\bth ,g\right)=[\bZ_{i} \left(t\right),\, g_{1} \left(W_{i} \left(t\right)\right),\, \left(g_{1} \left(W_{i} \left(t\right)\right)-\tau \right)_{+} ,\, \left(-\omega \right)\cdot {\rm I} \left\{g_{1} \left(W_{i} \left(t\right)\right)>\tau \right\}]$,
and define $S^{\left(0\right)}\left(t,\, \bth ,\, \, g\right)=\frac{1}{n} \sum _{i=1}^{n}Y_{i} \left(t\right)\,  \exp \left(\bps ^{T} \bV_{i} \left(g\left(t,\tau \right)\right)\right)$ , $s^{\left(0\right)}\left(t,\, \bth ,\, g\right)=E\left(S^{\left(0\right)}\left(t,\, \bth ,\, g\right)\right)\, \, $,
\\
$S^{\left(1\right)} \left(t,\bth ,g\right)=\frac{1}{n} \sum _{i=1}^{n}Y_{i} \left(t\right)\exp \left(\bps ^{T} \bV_{i} \left(g\left(t,\tau \right)\right)\right)\bxi _{i}  \left(t,\bth ,g\right)$.

\noindent Let  $\hat{\bps}$  and $\hat{\bth}$ denote the resulting estimators of $\bps $ and $\bth $, respectively.

\noindent Because of the discontinuity in the derivative of the log partial likelihood with respect to $\tau $, conventional maximization techniques cannot be applied. The obvious approach is to do a grid search over $\tau $, that is, maximizing the log partial likelihood over $\bps $ for a fixed value of $\tau $ (by taking $\frac{\partial l_{p}\left(\bth ,g\right)}{\partial \bps } =0$), this yields $\hat{\bps}$ , and then searching for $\tau$ such that $\left(\tau ,{\hat{\bps}}\right)$ maximizing the log likelihood, and this yields $\hat{\tau }$. In this work we use the bisection method over $\tau $ instead of a grid search, as we describe in the simulation chapter.


\noindent Define
\[\begin{array}{l} {\bQ\left(t^{*} ,\bth ,g\right)=E\left[\int _{0}^{t^{*} }\left(\bps ^{T} \bV\left(g\left(t,\tau \right)\right)-\log \left(s\left(t,\, \bth ,\, \, g\right)\right)\right)\bar{\tilde{\lambda }}\left(t\right)dt \right]} \\ {=E\left[\int _{0}^{t^{*} }\bps ^{T} \bV\left(g\left(t,\tau \right)\right)\bar{\tilde{\lambda }}\left(t\right)dt \right]-\int _{0}^{t^{*} }\log \left(s\left(t,\, \bth ,\, \, g\right)\right)\bar{\tilde{\lambda }}\left(t\right)dt \,} \end{array}\]
and define $\bth^*$ to be the solution of the equation $\bQ\left(t^{*},\bth,g\right)=0$.

\noindent Based on GA1-GA4 we have the following properties:

\noindent (i) The vector $\bV\left(g\left(t,\tau \right)\right)$ is bounded.

\noindent (ii) As in Andersen and Gill (1982) Theorem III.1, $\mathop{\sup }\limits_{\tsub,\bthsub } \left|S^{\left(0\right)}\left(t,\, \bth ,\, \, g\right)-s^{\left(0\right)}\left(t,\, \bth ,\, \, g\right)\right|\mathop{\mathop{\to }\limits^{a.s.} }\limits_{n\to \infty } 0$.

\noindent (iii) The log partial likelihood $l_{p} \left(\bth ,g\right)\, \, :\bTh \to R$ is Lipschitz.

\noindent (iv) The log partial likelihood $l_{p} \left(\bth ,g\right)\, \, :\bTh \to R$ is a continuous function in $\bth $.

\noindent (v) The log partial likelihood $l_{p} \left(\bth ,g\right)\, \, :\bTh \to R$ is a bounded function.

\noindent Using GA1-GA4 and the above properties (i)-(v), it can be shown that
$$
\mathop{\sup }\limits_{\bthsub \in \bThsub } \left|n^{-1} l_{p} \left(\bth ,g\right)-\bQ\left(t^{*} ,\bth ,g\right)\right|\mathop{\mathop{\to }\limits_{n\to \infty } }\limits^{p} 0
$$
that is, $\bQ\left(t^{*} ,\bth ,g\right)$ is the limit function of $n^{-1} l_{p} \left(\bth ,g\right)$.
(This can be proved using the process $n^{-1} \left(l_{p} \left(\bth ,g\right)-l_{p} \left(\bth ^{*} ,g\right)\right)$,
as in Andersen and Gill (1982)) and using Van der Vaart (1998) page 46). \\

We impose the following additional condition. \\

\noindent \textbf{Condition A.} For $t\in [0,t^{*} ]$, the matrix $\bSig \left(t^{*} ,\bth ,g\right)=-\frac{\partial ^{2} }{\partial \bth \partial \bth ^{T} } \bQ\left(t^{*} ,\bth ,g\right)$ is positive definite for \textbf{$\bth =\bth ^{*} $}.

\subsubsection{Heuristic consistency argument}
\noindent For a given $\tau$, the function $\bQ\left(t^{*} ,(\bpsi,\tau) ,g\right)$ is concave as a function
of $\bpsi$ and hence has a unique maximizer $\bpsi^*(\tau)$. The function $\tilde{\bQ}(t^*,\tau,g) = \bQ\left(t^{*},
(\bpsi^*(\tau),\tau) ,g\right)$ is generally not a concave function of $\tau$
and therefore it is difficult to
prove that it has a unique maximizer. But for the naive method,
plots of $\tilde{\bQ}(t^*,\tau,g)$ versus $\tau$ over a range of parameter settings suggest that
there is a unique maximizer $\tau^*$, leading to a unique maximizer $\bth^* = (\bpsi^*(\tau^*),\tau^*)$ of
$\bQ(t^*,\bth,g)$. Let us assume the existence of a unique maximizer $\bth^*$.
In view of the uniform convergence of $\ell_p$ to $\bQ$, the result of Foutz (1977) then implies that
there is some sequence of maximizers of $\ell_p$ that converges to $\bth^*$.
With the RC1 method, we observe some strange phenomena at the end of the changepoint
range when plotting the function  $\tilde{\bQ}(t^*,\tau,g)$, but usually there is a clear peak in the middle of the
range. Therefore, if we restrict the range of $\tau$, the argument we just made for the naive estimator
applies. In practice, we suggest restricting the range of $\tau$ to be between
the 5${}^{th}$ and the 95${}^{th}$ percentiles of the surrogate covariate $W\left(t\right)$.

\subsubsection{Asymptotic Normality}

Asymptotic normality is shown by an argument patterned after K\"{u}chenhoff and Wellisch (1997),
which is based on Huber (1967).
We define the likelihood score function as
\[\begin{array}{l} {\bU_{n} \left(\bV,\bth ,g\right)=\sum _{i=1}^{n}\delta _{i} \left[\bxi _{i} \left(T_{i}^{0} ,\bth ,g\right)-\frac{\sum _{j=1}^{n}Y_{j} \left(T_{i}^{0} \right)\exp \left(\bps ^{T} \bV_{j} \left(g\left(T_{i}^{0} ,\tau \right)\right)\right)\bxi _{j} \left(T_{i}^{0} ,\bth ,g\right) }{\sum _{j=1}^{n}Y_{j} \left(T_{i}^{0} \right)\exp \left(\bps ^{T} \bV_{j} \left(g\left(T_{i}^{0} ,\tau \right)\right)\right) } \right] } \\ {=\sum _{i=1}^{n}\delta _{i} \left[\bxi _{i} \left(T_{i}^{0} ,\bth ,g\right)-\frac{S^{\left(1\right)} \left(T_{i}^{0} ,\bth ,g\right)}{S^{\left(0\right)} \left(T_{i}^{0} ,\bth ,g\right)} \right] .} \end{array}\]
This function is equal to the gradient of $\ell_p(\bth,g)$ with respect to $\bth$ for all $\bpsi$ and all $\tau$ except when
$\tau$ is equal $g_1(W_i,\bZ_i)$ for some $i$, in which case $\ell_p$ is not diffentiable with respect to $\tau$.
Because of this exception, the MLE is not necessarily a solution of $\bU_{n} \left(\bV,{\hat{\bth} },g\right)=0$.
\\
\noindent Define $\bvarphi _{i} \left(\bV_{i} ,\bth ,g\right)=\delta _{i} \left[\bxi _{i} \left(T_{i}^{0} ,\bth ,g\right)-\frac{S^{\left(1\right)} \left(T_{i}^{0} ,\bth ,g\right)}{S^{\left(0\right)} \left(T_{i}^{0} ,\bth ,g\right)} \right]$,
${\tilde{\bvarphi}}_{i} \left(V_{i} ,\bth ,g\right)=\delta _{i} \left[\bxi _{i} \left(T_{i}^{0} ,\bth ,g\right)-\frac{s^{\left(1\right)} \left(T_{i}^{0} ,\bth ,g\right)}{s^{\left(0\right)} \left(T_{i}^{0} ,\bth ,g\right)} \right].$
Also define ${\btU}_{n} \left(V,\bth ,g\right)=\sum _{i=1}^{n}{\tilde{\bvarphi} }_{i} \left(\bV_{i} ,\bth ,g\right) $,
$\bPs _{n} \left(\bth \right)=\frac{1}{n} \sum _{i=1}^{n}\bvarphi _{i} \left(\bV_{i} ,\bth ,g\right),$
and $\tilde{\bPs }_{n} \left(\bth \right)=\frac{1}{n} \sum _{i=1}^{n}{\tilde{\bvarphi} }_{i} \left(\bV_{i} ,\bth ,g\right) $. Denote $\tilde{\bQ}\left(\bth \right)=E\left[{\tilde{\bvarphi} }_{i} \left(\bV_{i} ,\bth ,g\right)\right]$.

\noindent Under Conditions GA1-GA4 and Condition A, we have the following lemmas.

\noindent \textbf{Lemma 1.} ${\hat{\bth}}$ is an asymptotic solution of the score equations, that is,
\begin{equation}
n^{{-1 \mathord{\left/{\vphantom{-1 2}}\right.\kern-\nulldelimiterspace} 2} } \bU_{n} \left(\bV,{\hat{\bth}},g\right)\mathop{\mathop{\to }\limits_{} }\limits^{p} 0.
\label{CP}
\end{equation}

\noindent \textbf{Lemma 2. }

\noindent Define
$$
u\left(v,\bth ,d\right)=\mathop{\sup }\limits_{\left\| \alphsub -\bthsub \right\| \le \dsub} \left\| \bvarphi \left(v,\alpha
,g\right)-\bvarphi \left(v,\bth ,g\right)\right\|
$$
Then the following properties are satisfied (corresponding to (N-2) and (N-3) in Huber): \\
(N-2) There are strictly positive numbers $a,\, b,\, c,\, d_{0} $ such that

\noindent (i) $\left\| \tilde{\bQ}\left(\bth \right)\right\| \ge a\left\| \bth -\bth ^{*} \right\| $ for $\left\| \bth -\bth ^{*} \right\| \le d_{0} $.

\noindent (ii) $E\left(u\left(v,\bth ,d\right)\right)\le b\cdot d$ for $\left\| \bth -\bth ^{*} \right\| +d\le d_{0} $, $d\ge 0$

\noindent (iii) $E\left(u\left(v,\bth ,d\right)^{2} \right)\le c\cdot d$ for $\left\| \bth -\bth ^{*} \right\| +d\le d_{0} $, $d\ge 0$.

\noindent (N-3) The expectation $E\left(\left\| {\tilde{\bvarphi}}\left(v,\bth ^{*} ,g\right)\right\| ^{2} \right)$ is finite. \\

We now state our main results. \\

\noindent \textbf{Theorem} \textbf{1.}
\noindent Define $\tilde{h}_{i} \left(\bth ^{*} \right) \equiv \tilde{h}\left(\bV_{i} ,\bth ^{*} ,g\right)$ as follows:
\begin{align}
\tilde{h}_{i} \left(\bth ^{*} \right)
& =\frac{\delta _{i} }{s^{\left(0\right)} \left(T_{i}^{0} ,\bth ^{*} ,g\right)}
\frac{1}{n} \sum _{i=1}^{n}Y_{i} \left(T_{i}^{0} \right)\bxi _{i} \left(T_{i}^{0} ,\bth ^{*} ,g\right)
\exp \left(\bps ^{T} \bV_{i} \left(T_{i}^{0} ,\bth ^{*} ,g\right)\right)  \nonumber \\
& \hspace*{36pt}  -\frac{s^{\left(1\right)}\left(T_{i}^{0} ,\bth ^{*} ,g\right)}
{s^{\left(0\right)} \left(T_{i}^{0} ,\bth ^{*} ,g\right)}
\frac{1}{n} \sum _{i=1}^{n}Y_{i} \left(T_{i}^{0} \right)\exp \left(\bps ^{T} V_{i} \left(T_{i}^{0} ,\bth ^{*} ,g\right)\right)
\label{hi}
\end{align}
If $P\left(\| {\hat{\bth}}_{n} -\bth ^{*} \| \le d_{0} \right)\to 1$ and we define
$\tilde{\tilde{\bvarphi}}_{i} \left(\bth ^{*} \right)\equiv \tilde{\bvarphi}_{i} \left(\bth ^{*} \right)
+\tilde{h}_{i} \left(\bth ^{*} \right)$, then under Lemma 1 and Lemma 2,
$$
\sqrt{n} \tilde{\bQ}\left({\hat{\bth}}_{n} \right)
= \frac{1}{\sqrt{n} } \sum _{i=1}^{n}\tilde{\tilde{\bvarphi}}_{i} \left(\bth ^{*} \right) + o_p(1)
$$

\noindent \em{Remark}\rm: By definition, $E[\tilde{\bvarphi}_{i} \left(\bth ^{*} \right)]=\bzro$, and it is easily seen
that $E[\tilde{h}_{i} \left(\bth ^{*} \right)]=\bzro$. Thus $E[\tilde{\tilde{\bvarphi}}_{i} \left(\bth ^{*} \right)]=\bzro$. \\

\noindent \textbf{Corollary 1.} Under the conditions of Theorem 1, assume that $\bQ\left(\bth \right)$
has a non-singular derivative $\bLam $ at $\bth ^{*} $.
Then, $\sqrt{n} \left({\hat{\bth}}-\bth ^{*} \right)$ is asymptotically normal with mean zero
and covariance matrix $\Lambda ^{-1} \bC\left(\bth ^{*} \right)\left(\bLam ^{-1} \right)^{T} $,
where $\bC\left(\bth ^{*} \right)$ is the covariance matrix of
$\tilde{\tilde{\bvarphi}}_{i} \left(\bth ^{*} \right) $. \\

\noindent By the strong law of large numbers, $\bC\left(\bth \right)$ can be estimated by
$\hat{\bC}\left(\bth \right)
= n^{-1} \sum _{i=1}^{n}\tilde{\tilde{\bPs }}_{i} \left(\bth \right)^{\otimes 2}$, where
$\tilde{\tilde{\bPs }}_{i} \left(\bth \right)=\tilde{\tilde{\bPs }}_{1,i} \left(\bth \right)
+ \tilde{\tilde{\bPs }}_{2,i} \left(\bth \right)$, with $\tilde{\tilde{\bPs }}_{1,i} \left(\bth \right)=\delta _{i} \left[\bxi _{i} \left(T_{i}^{0} ,\bth ,g\right)-\frac{S^{\left(1\right)} \left(T_{i}^{0} ,\bth ,g\right)}{S^{\left(0\right)} \left(T_{i}^{0} ,\bth ^{*} ,g\right)} \right]$ and
$\tilde{\tilde{\bPs }}_{2,i} \left(\bth \right)=\frac{1}{S^{\left(0\right)} \left(T_{i}^{0} ,\bth ,g\right)}
\left[\begin{array}{l} {Y_{i} \left(T_{i}^{0} \right)\bxi _{i} \left(T_{i}^{0} ,\bth ,g\right)\exp \left(\bps ^{T} \bV_{i} \left(g\left(T_{i}^{0} ,\tau \right)\right)\right)} {-\frac{S^{\left(1\right)} \left(T_{i}^{0} ,\bth ,g\right)}{S^{\left(0\right)} \left(T_{i}^{0} ,\bth ,g\right)} Y_{i} \left(T_{i}^{0} \right)\exp \left(\bps ^{T} \bV_{i} \left(g\left(T_{i}^{0} ,\tau \right)\right)\right)}
\end{array}\right].$

\noindent Since ${\hat{\bth}}_{n}$ is consistent and $\bC$ is continuous, we can estimate
$\bC\left(\bth ^{*} \right)$ consistently by $\hat{\bC}\left({\hat{\bth}}_{n} \right)$.

\subsection{RC2 method}
\noindent The RC2 estimator involves replacing of $X_{i} \left(t\right)$ in the term $\beta X_{i} \left(t\right)$ by a surrogate $g_{1} \left(w,z\right)=\mu \left(w,z\right)$ and $\left(X_{i} \left(t\right)-\tau \right)_{+} $ in the term $\omega \left(X_{i} \left(t\right)-\tau \right)_{+} $ by a surrogate
\\
 $g_{2} \left(w,z\right)=E\left[\left. \left(x-\tau \right)_{+} \right|W=w\,,\,\bZ=z\right]$.
The log partial likelihood is then
\[l_{p} \left(t^{*} ,\bth ,g\right)=\sum _{i=1}^{n}\int _{0}^{t^{*} }\bth ^{T} \bV_{i} \left(g\left(t\right)\right)dN_{i} \left(t\right) -\int _{0}^{t^{*} }\log  \left[\sum _{j=1}^{n}Y_{j} \left(t\right)\exp \left(\bth ^{T} \bV_{i} \left(g\left(t\right)\right)\right) \right]d\bar{N}\left(t\right) \, . \]
Let $\bV_{1} \left(g\left(t,\tau \right)\right)$ denote a vector of length $p+3$ in which the first $p$ components are the elements of $\bZ\left(t\right)$, the $\left(p+1\right)$-th component is $\mu \left(w,z\right)$, the $\left(p+2\right)$-th component is
\\
$E\left[\left. \left(x-\tau \right)_{+} \right|W=w\, ,\, \bZ=z\right]$, and the $\left(p+3\right)$-th component is $\omega \frac{\partial E\left[\left. \left(x-\tau \right)_{+} \right|W=w\, ,\, \bZ=z\right]}{\partial \tau } $,

\noindent where  $\frac{\partial E\left[\left. \left(X_{i} -\tau \right)_{+} \right|W_{i} ,\bZ_{i} \right]}{\partial \tau } =-\left(1-\Phi \left(\frac{-\mu \left(W_{i} ,\bZ_{i} \right)+\tau }{\eta } \right)\right)\, \, \,$. Define (with $a^{\otimes 2} $ for a vector $a$ defined as $aa^{T} $) $S^{\left(0\right)} \left(t,\, \bth ,\, \, g\right)=\frac{1}{n} \sum _{i=1}^{n}Y_{i} \left(t\right)\,  \exp \left(\bth ^{T} \bV_{i} \left(g\left(t\right)\right)\right)$,
 \\
 $S^{\left(1\right)} \left(t,\, \bth ,\, g\right)=\frac{1}{n} \sum _{i=1}^{n}Y_{i} \left(t\right)\,  \bV_{1,i} \left(g\left(t\right)\right)\exp \left(\bth ^{T} \bV_{i} \left(g\left(t\right)\right)\right)$,
 \\
 $S^{\left(2\right)} \left(t,\, \bth ,\, g\right)=\frac{1}{n} \sum _{i=1}^{n}Y_{i} \left(t\right)\,  \bV_{1,i} \left(g\left(t\right)\right)^{\otimes 2} \exp \left(\bth ^{T} \bV_{i} \left(g\left(t\right)\right)\right)$.

\noindent The score function is $\bU\left(t^{*} ,\, \bth ,\, g\right)=\sum _{i=1}^{n}\int _{0}^{t^{*} }\left[\bV_{1,i} \left(g\left(t\right)\right)-\frac{S^{\left(1\right)} \left(t,\, \bth ,\, g\right)}{S^{\left(0\right)} \left(t,\, \bth ,\, g\right)} \right]  dN_{i} \left(t\right)$.

\noindent Let $\hat{\bth }$ denote the resulting estimator.
Define
\[\bQ\left(t^{*},\bth,g\right)\ ={\int _{0}^{t^{*} }\tilde{s}^{\left(1\right)} \left(t,\bth ,g\right)dt-\int _{0}^{t^{*} }\frac{s^{\left(1\right)} \left(t^{*},\bth ,g\right)}{s^{\left(0\right)} \left(t,\bth,g\right)} \tilde{s}^{\left(0\right)} \left(t,\bth\right) dt }.\]
  and denote by $\bth ^{*} $ the solution of  $\bQ\left(t^{*} ,\bth ,g\right)\, =0$.

 Additional notations are as in Agami et al. (2018) Appendix A.1, where $\bV_{i}$ is replaced with $\bV_{1,i}$ that was defined above.

\noindent \textbf{Proposition 1}. Assume that GA1-GA3 hold and that
$\bSig \left(t^{*} ,\bth ,g\right)$ is positive definite at \textbf{$\bth =\bth ^{*} $}.
Then

\noindent  (i) For $n$ sufficiently large, the estimator $\hat{\bth }$ is the {unique} solution of $\bU\left(t^{*} ,\bth ,g\right)=0$.

\noindent (ii) The estimator $\hat{\bth }$ is a {consistent} estimator of $\bth ^{*} $.

\noindent \textbf{Proof of Proposition 1. }

\noindent The proof is similar to the proposition's proof in Agami et. al (2018) Chapter 3.1.
\noindent \textbf{}

\noindent \textbf{}

\noindent \textbf{Proposition 2}:\textbf{ Asymptotic Normality\underbar{}}

\noindent $n^{{1 \mathord{\left/{\vphantom{1 2}}\right.\kern-\nulldelimiterspace} 2} } \left(\hat{\bth }-\bth ^{*} \right)$ convergences in distribution to a mean-zero multivariate normal distribution whose covariance matrix can be consistently estimated by $\bomg \left(t^{*} ,\, \hat{\bth },\, g\right)=n^{-1} I\left(t^{*} ,\, \hat{\bth },\, g\right)^{-1} \hat{A}\left(t^{*} ,\, \hat{\bth },\, g\right)n^{-1} I\left(t^{*} ,\, \hat{\bth },\, g\right)^{-1} $, where $\hat{A}\left(t^{*} ,\, \bth ,\, g\right)=\frac{1}{n} \sum _{i=1}^{n}\hat{H}_{i}  \left(t^{*} ,\, \bth ,\, g\right)^{\otimes 2} $ and
\\
$\begin{array}{l} {\hat{H}_{i} \left(t^{*} ,\, \bth ,\, g\right)} \\ {=\int _{0}^{t^{*} }\left\{\bV_{1,i} \left(g\left(t\right)\right)-\frac{S^{\left(1\right)} \left(t,\, \bth ,\, g\right)}{S^{\left(0\right)} \left(t,\, \bth ,\, g\right)} \right\} dN_{i} \left(t\right)-\int _{0}^{t^{*} }\frac{Y_{i} \left(t\right)\exp \left(\bth ^{T} \bV_{i} \left(g\left(t\right)\right)\right)}{S^{\left(0\right)} \left(t,\, \bth ,\, g\right)} \left\{\bV_{1,i} \left(g\left(t\right)\right)-\frac{S^{\left(1\right)} \left(t,\, \bth ,\, g\right)}{S^{\left(0\right)} \left(t,\, \bth ,\, g\right)} \right\} d\tilde{F}_{n} \left(t\right)\, \,.} \end{array}$\textbf{}

\noindent \textbf{Proof of Proposition 2. }

\noindent The proof is similar to the to the proposition's proof in Agami et. al (2018) Chapter 3.1,
where $n^{-1/2} \bU\left(t^{*} ,\bth ^{*} ,g\right)$ is asymptotically  equivalent to  $n^{-1/2} \sum h_{i} \left(t^{*} ,\bth ^{*} ,g\right) $, with

\noindent $h_{i} \left(t^{*} ,\bth ,g\right)
\\
=\int _{0}^{t^{*} }\left\{\bV_{1,i} \left(g\left(t\right)\right)-\frac{s^{\left(1\right)} \left(t,\bth ,g\right)}{s^{\left(0\right)} \left(t,\bth ,g\right)} \right\} dN_{i} \left(t\right)-\int _{0}^{t^{*} }\frac{Y_{i} \left(t\right)\exp \left\{\bth ^{T} \bV_{i} \left(g\left(t\right)\right)\right\}}{s^{\left(0\right)} \left(t,\bth ,g\right)}  \left\{\bV_{1,i} \left(g\left(t\right)\right)-\frac{s^{\left(1\right)} \left(t,\bth ,g\right)}{s^{\left(0\right)} \left(t,\bth ,g\right)} \right\}d\tilde{F}\left(t\right)\,.$ $\square$

\subsection{RR method}
\noindent The RR involves replacing $\exp \left(\beta X_{i} \left(t\right)+\omega \left(X_{i} \left(t\right)-\tau \right)_{+} \right)$ with \\$E\left(\left. \exp \left(\beta X_{i} \left(t\right)+\omega \left(X_{i} \left(t\right)-\tau \right)_{+} \right)\right|W_{i} \left(t\right),\, \bZ_{i} \left(t\right)\right)$. The development of asymptotic theory in the RR method with unknown changepoint is the same as in the case of known changepoint, but with $\bth$ that includes $\tau$. Therefore, using the above definition of $\bV\left(t\right)$, we have the same results as in the case of known changepoint, see Agami et. al (2018) Chapter 3.2.

\subsection{MPPLE}
\noindent let $\bV_{i} \left(t\right)$ denote a vector of length $p+2$ in which the first $p$ components are the elements of $\bZ_{i} \left(t\right)$, the $\left(p+1\right)-$th component is $X_{i} \left(t\right)$, and the $\left(p+2\right)-$th component is $\left(X_{i} \left(t\right)-\tau \right)_{+} $. In addition, denote $\bth =\left(\gamma ^{T} ,\beta ,\omega ,\tau \right)$, and denote the true values of $\bth $ by $\bth _{0} $.

\noindent The normalized log likelihood function is:
\[l_{p} \left(\bth \right)=\frac{1}{n} \sum _{i=1}^{n}\delta _{i}  \left[\tilde{\phi }\left(\bth ,\bV_{i} \left(T_{i}^{0} \right),\Lambda _{0} \left(T_{i}^{0} \right)\right)-\log \sum _{j=1}^{n}Y_{j} \left(T_{i}^{0} \right)\exp \left(\tilde{\phi }\left(\bth ,\bV_{i} \left(T_{i}^{0} \right),\Lambda _{0} \left(T_{i}^{0} \right)\right)\right) \right],\]
where
\[\begin{array}{l} {\tilde{\phi }\left(\bth ,\bV_{i} \left(T_{i}^{0} \right),\Lambda _{0} \left(T_{i}^{0} \right)\right)}\\ {=\log \int _{-\infty }^{\infty }\left\{\begin{array}{l} {\exp \left(-\Lambda _{0} \left(T_{i}^{0} \right)\cdot \exp \left(\bgam ^{T} \bZ_{i} \left(T_{i}^{0} \right)+\beta X_{i} \left(T_{i}^{0} \right)+\omega \left(X_{i} \left(T_{i}^{0} \right)-\tau \right)_{+} \right)\right)} \\ {\times \exp \left(\bgam ^{T} \bZ_{i} \left(T_{i}^{0} \right)+\beta X_{i} \left(T_{i}^{0} \right)+\omega \left(X_{i} \left(T_{i}^{0} \right)-\tau \right)_{+} \right)} \end{array}\right\}f_{\left. X\right|W,\bZ} \left(x\right)dx } \\ {-\log \int _{-\infty }^{\infty }\left\{\exp \left(-\Lambda _{0} \left(T_{i}^{0} \right)\cdot \exp \left(\bgam ^{T} \bZ_{i} \left(T_{i}^{0} \right)+\beta X_{i} \left(T_{i}^{0} \right)+\omega \left(X_{i} \left(T_{i}^{0} \right)-\tau \right)_{+} \right)\right)\right\}f_{\left. X\right|W,\bZ} \left(x\right)dx } \\ {\equiv \left({\rm I} \right)-\left({\rm I} {\rm I} \right)\, .} \end{array}\]
We can write $\left({\rm I} \right)$ and $\left({\rm I} {\rm I} \right)$ as follows.
\[\begin{array}{l} {\left({\rm I} \right)=} \\ {\log \left[\begin{array}{l} {\int _{-\infty }^{\tau }\left\{\exp \left(-\Lambda _{0} \left(T_{i}^{0} \right)\cdot \exp \left(\bgam ^{T} \bZ_{i} \left(T_{i}^{0} \right)+\beta X_{i} \left(T_{i}^{0} \right)\right)\right) \exp \left(\bgam ^{T} \bZ_{i} \left(T_{i}^{0} \right)+\beta X_{i} \left(T_{i}^{0} \right)\right)\right\}f_{\left. X\right|W,\bZ} \left(x\right)dx } \\ {+\int _{\tau }^{\infty }\left\{\begin{array}{l} {\exp \left(-\Lambda _{0} \left(T_{i}^{0} \right)\cdot \exp \left(\bgam ^{T} \bZ_{i} \left(T_{i}^{0} \right)+\beta X_{i} \left(T_{i}^{0} \right)+\omega \left(X_{i} \left(T_{i}^{0} \right)-\tau \right)\right)\right)} \\ {\times \exp \left(\bgam ^{T} \bZ_{i} \left(T_{i}^{0} \right)+\beta X_{i} \left(T_{i}^{0} \right)+\omega \left(X_{i} \left(T_{i}^{0} \right)-\tau \right)\right)} \end{array}\right\} f_{\left. X\right|W,\bZ} \left(x\right)dx} \end{array}\right]} \end{array}\]
Denote $\psi _{1} \left(v;\bth \right)= \exp \left(\bgam ^{T} z+\beta x\right)$, and $\psi _{2} \left(v;\bth \right)= \exp \left(\gamma ^{T} z+\beta x+\omega \left(x-\tau \right)\right)$. Then, using these notations we can write that  $\tilde{\phi }\left(\bth ,v,c \right)= \left({\rm I} \right)+\left({\rm I} {\rm I} \right)$, where
\[\begin{array}{l} {\left({\rm I} \right)=\log \left[\int _{-\infty }^{\tau }\left\{e^{-c\psi _{1} \left(v;\bth \right)} \psi _{1} \left(v;\bth \right)\right\}f_{\left. X\right|W,\bZ} \left(x\right)dx+\int _{\tau }^{\infty }\left\{e^{-c\psi _{2} \left(v;\bth \right)} \psi _{2} \left(v;\bth \right)\right\} f_{\left. X\right|W,\bZ} \left(x\right)dx \right]} \\ {\left({\rm I} {\rm I} \right)=\log \left[\int _{-\infty }^{\tau }\left\{e^{-c\left(v;\bth \right)} \right\}f_{\left. X\right|W,\bZ} \left(x\right)dx+\int _{\tau }^{\infty }\left\{e^{-c\psi _{2} \left(v;\bth \right)} \right\} f_{\left. X\right|W,\bZ} \left(x\right)dx \right]\, .} \end{array}\]
\textbf{Conditions:}

\noindent A. $\bth _{0} $ is an interior point.

\noindent B. The matrix of the second derivatives of the limit log likelihood function respect to $\bth $ is positive definite for $\bth =\bth _{0} $, and semipositive definite for all \textbf{$\bth $}.

\noindent

\noindent Assumptions I.-IX. in Zucker (2005) are the same as the assumptions in our setting, except of the assumption of twice continuously differentiable of the function $\psi \left(x;\bth \right)=\exp \left(\bth ^{T} x\right)$ with respect to $\bth $ over $\bTh $. We have instead the assumption of twice continuously differentiable of the functions  $\psi _{1} \left(v;\bth \right)$ and $\psi _{2} \left(v;\bth \right)$ with respect to $\bth $ over $\bTh $.

\noindent By assumptions GA2 (the third part of this assumption) and GA3, and because $f_{\left. X\right|W,\bZ} \left(x\right)$ is bounded by some constant, the derivative of the function inside of each integral is bounded. Then we can interchange derivative with integral (Bartle (1966) Corollary 5.9) and write that :
\[\begin{array}{l} {\frac{\partial \left({\rm I} \right)}{\partial \bxi } =} \\ {\, \, \, \, \left[\begin{array}{l} {\int _{-\infty }^{\tau }\left\{e^{-c\psi \left(v;\bth \right)} \left(-c\right)\left(\frac{\partial \psi _{1} \left(v;\bth \right)}{\partial \bxi } \right)\psi _{1} \left(v;\bth \right)+e^{-c\psi _{1} \left(v;\bth \right)} \left(\frac{\partial \psi _{1} \left(v;\bth \right)}{\partial \bxi } \right)\right\}f_{\left. X\right|W,Z} \left(x\right)dx } \\ {+\int _{\tau }^{\infty }\left\{e^{-c\psi _{2} \left(v;\bth \right)} \left(-c\right)\left(\frac{\partial \psi _{2} \left(v;\bth \right)}{\partial \bxi } \right)\psi _{2} \left(v;\bth \right)+e^{-c\psi _{2} \left(v;\bth \right)} \left(\frac{\partial \psi _{2} \left(v;\bth \right)}{\partial \bxi } \right)\right\} f_{\left. X\right|W,\bZ} \left(x\right)dx} \end{array}\right]} \\ {\, \, \, \, \, \, \, \, \, \times \frac{1}{\left[\int _{-\infty }^{\tau }\left\{e^{-c\psi _{1} \left(v;\bth \right)} \psi _{1} \left(v;\bth \right)\right\}f_{\left. X\right|W,\bZ} \left(x\right)dx+\int _{\tau }^{\infty }\left\{e^{-c\psi _{2} \left(v;\bth \right)} \psi _{2} \left(v;\bth \right)\right\} f_{\left. X\right|W,\bZ} \left(x\right)dx \right]} } \\ {\frac{\partial \left({\rm I} {\rm I} \right)}{\partial \bxi } =} \\ {\frac{\int _{-\infty }^{\tau }\left\{e^{-c\psi _{1} \left(v;\bth \right)} \left(-c\right)\left(\frac{\partial \psi _{1} \left(v;\bth \right)}{\partial \bxi } \right)\right\}f_{\left. X\right|W,Z} \left(x\right)dx+\int _{\tau }^{\infty }\left\{e^{-c\psi _{2} \left(v;\bth \right)} \left(-c\right)\left(\frac{\partial \psi _{2} \left(v;\bth \right)}{\partial \bxi } \right)\right\} f_{\left. X\right|W,\bZ} \left(x\right)dx }{\left[\int _{-\infty }^{\tau }\left\{e^{-c\psi _{1} \left(v;\bth \right)} \right\}f_{\left. X\right|W,\bZ} \left(x\right)dx+\int _{\tau }^{\infty }\left\{e^{-c\psi _{2} \left(v;\bth \right)} \right\} f_{\left. X\right|W,\bZ} \left(x\right)dx \right]} } \end{array}\]
where
\[\frac{\partial \psi _{1} \left(v;\bth \right)}{\partial \bgam } =\exp \left(\bgam ^{T} z+\beta x\right)\cdot z, \frac{\partial \psi _{2} \left(v;\bth \right)}{\partial \bgam } =\exp \left(\bgam ^{T} z+\beta x+\omega \left(x-\tau \right)\right)\cdot z,\]
\[\frac{\partial \psi _{1} \left(v;\bth \right)}{\partial \beta } =\exp \left(\bgam ^{T} z+\beta x\right)\cdot x, \frac{\partial \psi _{2} \left(v;\bth \right)}{\partial \beta } =\exp \left(\bgam ^{T} z+\beta x+\omega \left(x-\tau \right)\right)\cdot x,\]
\[\frac{\partial \psi _{1} \left(v;\bth \right)}{\partial \omega } =0, \frac{\partial \psi _{2} \left(v;\bth \right)}{\partial \omega } =\exp \left(\bgam ^{T} z+\beta x+\omega \left(x-\tau \right)\right)\cdot \left(x-\tau \right).\]
In addition, we have for the first integral in $\left({\rm I} \right)$ and $\left({\rm I} {\rm I} \right)$ that $\frac{\partial }{\partial \tau } \int _{-\infty }^{\tau }\left\{e^{-c\psi _{1} \left(v;\bth \right)} \psi _{1} \left(v;\bth \right)\right\}f_{\left. X\right|W,\bZ} \left(x\right)dx$
 \\
 $=\left\{\exp \left(-c\cdot \exp \left(\bgam ^{T} z+\beta \cdot \tau \right)\right)\cdot \exp \left(\bgam ^{T} z+\beta \cdot \tau \right)\right\}f_{\left. X\right|W,\bZ} \left(x\right)\frac{\partial }{\partial \tau } \int _{-\infty }^{\tau }\left\{e^{-c\psi _{1} \left(v;\bth \right)} \right\}f_{\left. X\right|W,\bZ} \left(x\right)dx= \left\{\exp \left(-c\cdot \exp \left(\bgam ^{T} z+\beta \cdot \tau \right)\right)\right\}f_{\left. X\right|W,\bZ} \left(x\right)\, .$
For the second integral of $\left({\rm I} \right)$ and $\left({\rm I} {\rm I} \right)$, since the expressions inside each integral are continuous functions in $\tau $ (as a product or a composition of continuous functions), and since their derivatives are continuous functions in $\tau $, then we can use the Leibnitz's Rule (Kaplan (2002), Chapter 4) and write that
\[\begin{array}{l} {\frac{\partial }{\partial \tau } \int _{-\infty }^{\tau }\left\{e^{-c\psi _{2} \left(v;\theta \right)} \psi _{2} \left(v;\bth \right)\right\}f_{\left. X\right|W,\bZ} \left(x\right)dx= \left. e^{-c\psi _{2} \left(v;\bth \right)} \psi _{2} \left(t\right)\right|_{x=\tau } } \\ {+\int _{-\infty }^{\tau }\frac{\partial }{\partial \tau } \left\{e^{-c\psi _{2} \left(v;\theta \right)} \psi _{2} \left(v;\bth \right)\right\}f_{\left. X\right|W,\bZ} \left(x\right)dx }, \end{array}\]
 $\frac{\partial }{\partial \tau } \int _{-\infty }^{\tau }\left\{e^{-c\psi _{2} \left(v;\bth \right)} \right\}f_{\left. X\right|W,\bZ} \left(x\right)dx= \left. e^{-c\psi _{2} \left(v;\bth \right)} \right|_{x=\tau } +\int _{-\infty }^{\tau }\frac{\partial }{\partial \tau } \left\{e^{-c\psi _{2} \left(v;\bth \right)} \right\}f_{\left. X\right|W,\bZ} \left(x\right)dx $

\noindent where
\[\frac{\partial }{\partial \tau } \left\{e^{-c\psi _{2} \left(v;\bth \right)} \psi _{2} \left(v;\bth \right)\right\}=e^{-c\psi _{2} \left(v;\bth \right)} \left(-c\right)\left(\frac{\partial \psi _{2} \left(v;\bth \right)}{\partial \tau } \right)\psi _{2} \left(v;\bth \right)+e^{-c\psi _{2} \left(v;\bth \right)} \left(\frac{\partial \psi _{2} \left(v;\bth \right)}{\partial \tau } \right),\]
\[\frac{\partial }{\partial \tau } \left\{e^{-c\psi _{2} \left(v;\bth \right)} \right\}=e^{-c\psi _{2} \left(v;\bth \right)} \left(-c\right)\left(\frac{\partial \psi _{2} \left(v;\bth \right)}{\partial \tau } \right),\, \, \frac{\partial \psi _{2} \left(v;\bth \right)}{\partial \tau } =\exp \left(\bgam ^{T} z+\beta x+\omega \left(x-\tau \right)\right)\left(-\omega \right).\]

\noindent Then we use these derivatives in:
\[\begin{array}{l} {\frac{\partial \left({\rm I} \right)}{\partial \tau } =} \\ {\left[\begin{array}{l} {\int _{-\infty }^{\tau }\left\{e^{-c\psi _{1} \left(v;\bth \right)} \left(-c\right)\left(\frac{\partial \psi _{1} \left(v;\bth \right)}{\partial \tau } \right)\psi _{1} \left(v;\bth \right)+e^{-c\psi _{1} \left(v;\bth \right)} \left(\frac{\partial \psi _{1} \left(v;\bth \right)}{\partial \tau } \right)\right\}f_{\left. X\right|W,\bZ} \left(x\right)dx } \\ {\, \, \, +\int _{\tau }^{\infty }\left\{e^{-c\psi _{2} \left(v;\bth \right)} \left(-c\right)\left(\frac{\partial \psi _{2} \left(v;\bth \right)}{\partial \tau } \right)\psi _{2} \left(v;\bth \right)+e^{-c\psi _{2} \left(v;\bth \right)} \left(\frac{\partial \psi _{2} \left(v;\bth \right)}{\partial \tau } \right)\right\} f_{\left. X\right|W,\bZ} \left(x\right)dx} \end{array}\right]} \\ {\, \, \, \times \frac{1}{\left[\int _{-\infty }^{\tau }\left\{e^{-c\psi _{1} \left(v;\bth \right)} \psi _{1} \left(v;\bth \right)\right\}f_{\left. X\right|W,\bZ} \left(x\right)dx+\int _{\tau }^{\infty }\left\{e^{-c\psi _{2} \left(v;\bth \right)} \psi _{2} \left(v;\bth \right)\right\} f_{\left. X\right|W,\bZ} \left(x\right)dx \right]} } \\ {\frac{\partial \left({\rm I} {\rm I} \right)}{\partial \bxi } =} \\ {\frac{\int _{-\infty }^{\tau }\left\{e^{-c\psi _{1} \left(v;\bth \right)} \left(-c\right)\left(\frac{\partial \psi _{1} \left(v;\bth \right)}{\partial \tau } \right)\right\}f_{\left. X\right|W,\bZ} \left(x\right)dx+\int _{\tau }^{\infty }\left\{e^{-c\psi _{2} \left(v;\bth \right)} \left(-c\right)\left(\frac{\partial \psi _{2} \left(v;\bth \right)}{\partial \tau } \right)\right\} f_{\left. X\right|W,Z} \left(x\right)dx }{\left[\int _{-\infty }^{\tau }\left\{e^{-c\psi _{1} \left(v;\bth \right)} \right\}f_{\left. X\right|W,\bZ} \left(x\right)dx+\int _{\tau }^{\infty }\left\{e^{-c\psi _{2} \left(v;\bth \right)} \right\} f_{\left. X\right|W,\bZ} \left(x\right)dx \right]} \, .} \end{array}\]
The proof of consistency in Zucker (2005) is the same in our setting, where the assumptions of differentiability of $l_{p} \left(\bth \right)$ with respect to $\bth $ and the existence of continuity on $\bTh$ of the second order derivatives of $l_{p} \left(\bth \right)$ respect to $\bth $ are fulfilled in our setting. Therefore, ${\hat{\bth}}\mathop{\to }\limits^{p} \bth _{0} $.

\noindent Also, the proof of asymptotic normality of ${\hat{\bth}}$ in Zucker (2005) is the same in our setting. Therefore, $\sqrt{n} \left({\hat{\bth}}-\bth _{0} \right)\sim N\left(0,\bV\right)$, where $V$ is as in equation (13) in Zucker (2005).

\noindent
\subsection{Asymptotic Bias }
As noted earlier, the asymptotic limits of the naive, RC1 and RC2 estimators are the solution $\bth ^{*}$ of $\bQ \left(t,\bth ,g\right)=0$, where $\bQ \left(t,\bth,g\right)$ is the limit of $U\left(t,\bth,g\right)$ as \textit{n} tends to infinity. Similarly, the asymptotic limit of the RR1 estimator is the solution $\bth ^{*}$ of $\bQ\left(t,\bth\right)=0$, where $\bQ \left(t,\bth\right)$ is the limit of $U\left(t,\bth\right)$ as \textit{n} tends to infinity. The asymptotic bias is then $\bth ^{*}-\bth _{0}$. Hughes (1993) previously used this approach to evaluate the asymptotic bias of the naive estimator in the Cox model without a threshold effect.

\indent We computed the limiting values numerically for the naive,RC1, RC2, and RR1 estimators under the rare disease scenario where $n=50,000$ and cumulative incidence = 0.01. The calculations involved the Newton-Raphson method to find the points where the score function equals zero, we compared the results with those obtained in the simulation studies, for the case where the measurement error parameters are known. The results are detailed in the supplement. Both the theoretical and empirical bias are based on a model with one covariate and true parameters of  $\beta =\log \left(1.5\right)=0.405$ and $\omega =\log \left(2\right)=0.693$. The starting values for the Newton Raphson calculation in all methods were (0,0). Tables S.1. in the supplement present the results, where the asymptotic bias labeled by {\textit{theoretical}}, and the simulation results are labeled by {\textit{empirical}}. The variable {\textit{pct}} denotes the convergence percent over 1000 replications. Also we define the variable DELTA to be the difference between the theoretical result and the simulation result. Generally, the RR1 method has the least bias, typically negligible, except at the lower extreme values of $\tau$, where the relative bias is $\pm0.05$ for $\rho_{xw}$=0.8 and becomes larger as $\rho_{xw}$ decreases. Regrading the comparison with the simulation results: for the naive method, when $n=50,000$ and cumulative incidence = 0.01, theoretical and simulation results agreed closely, as expected. For the RC1 and RC2 methods, the agreement between the theoretical and simulation results was better for $\tau<0$ with n=200,000 (keeping cumulative incidence of 0.01)where for $\tau>0$, the agreement is similar with $n=50,000$ and $n=200,000$. The results were close, except at the lower extreme values of $\tau$ in which case this difference was large. Regarding the RR1 method, the results were close, except at the lower extreme values of $\tau$ in which case this difference was large. Consequently, there is a good agreement between the empirical and the theoretical results, so that generally one can use the theoretical results to evaluate the bias for a given scenario.

\section{Simulation Study}
In this section, we compare via a simulation study the various methods under several scenarios. and tables S.2. and S.3. in the supplement materials present the results. As a benchmark, we also present the estimates under the case of no measurement error.
The simulation design is the same as in Agami et al. (2018).
\\
 \\
\begin{landscape}
\begin{center}
\fontsize{7.35}{0.25}\selectfont

\begin{tabular}{  m{1.1cm}  m{1.1cm} m{0.8cm} m{1cm} m{1.3cm} m{1.3cm} m{1.1cm} m{1.2cm} m{1.1cm} m{1.2cm} m{1.1cm} m{1.2cm} m{0.9cm} m{1.25cm} m{1.05cm}  }
\mc{10}{c}{Table 1: Finite Sample Bias$^a$ in $\beta$, ($\beta$, $\omega$)=(0.405, 0.693)} \\
$\tau$ &Disease& $\rho _{xw}$& Naive & RC1(kn)$^b$& RC1(ukn)$^c$ & RC2(kn)$^b$ & RC2(ukn)$^c$ & RR1(kn)$^b$ & RR1(ukn)$^c$ & RR2(kn)$^b$ &
RR2(ukn)$^c$& SIMEX &MPPLE(kn)$^b$&MPPLE(ukn)$^c$ \\
\\
$\Phi^{-1}(0.1)$ &Common & 1  & 0.620 &  & &  &  & &  &  &  &  & &  \\
 \\
 & & 0.8  & 0.482&	1.313&	1.302&	1.319&	1.290&	1.273&	1.227&	0.992&	0.940&	0.714&	1.297&	\\
  \\
  & & 0.6  & -0.195&	1.231&	1.190&	1.256&	1.214&	1.279&	1.182&	0.634&	0.580&	-0.053&	1.499&	 \\
  \\
  & & 0.4  & -0.659&	1.128&	-0.643&	1.260&	1.144&	1.267&	0.915&	0.254&	0.048&	-0.602&	1.445&	 \\
  \\
  &Rare & 1  &*0.839$^d$ &  &  &  &  &  &  &  &  &  & &   \\
  \\
  & & 0.8  & 0.604&	1.233&	1.178&	1.307&	1.408&	0.632&	0.573&	-0.790&	-0.844 &  & &   \\
  \\
   & & 0.6  & -0.049&	-0.052&	0.301&	1.545&	1.445&	-0.436&	-0.822&	-2.322&	-2.624 &  & &   \\
  \\
   & & 0.4  &-0.573&	-3.086&	0.297&	1.194&	1.361&	-1.590&	0.501&	-3.503&	-0.920 &  & &   \\
  \\
 $\Phi^{-1}(0.25)$ &Common & 1  & -0.017 &  &  &  &  &  &  &  &  &  & &   \\
 \\
 & & 0.8  & -0.014&	0.537&	0.516&	0.228&	0.081&	0.122&	-0.026&0.053&	-0.067&	0.252&	0.260&	0.254\\
   \\
  & & 0.6  & -0.332&	0.851&	0.757&	0.793&	0.668&	0.626&	0.487&	0.219&	0.120&	-0.209&	0.923&	0.906   \\
  \\
  & & 0.4  & -0.699&	0.879&	0.822&	0.823&	0.744&	0.658&	0.384&	-0.090&	-0.285&	-0.646&	0.910&	1.081  \\
  \\
  &Rare & 1  &0.161 &  &  &  &  & &  &  &  &  & &   \\
  \\
  & & 0.8  & 0.219&	0.659&	0.658&	0.655&	0.647&	0.395&	0.413&	-0.084&	-0.096 &  & &  \\
  \\
   & & 0.6  & -0.178&	1.089&	1.021&	1.005&	1.032&	0.580&	0.396&	-0.536&	-0.627 &  & &  \\
  \\
   & & 0.4  & -0.605&	1.318&	*0.951&	1.251&	1.161&	-0.198&	0.280&	-1.790&	-1.118&  & & \\
  \\
 $\Phi^{-1}(0.5)$ &Common & 1  & -0.015 &  &  &  & &  & &  &  & & &   \\
 \\
 & & 0.8  & -0.261&	0.153&	0.153&	-0.049&	-0.083&	-0.033&	-0.070&	0.071&	0.043&	-0.127&	0.011&	0.003   \\
  \\
  & & 0.6 & -0.522&	0.324&	0.261&	0.078&	-0.030&	-0.018&	-0.066&	-0.067&-0.130&	-0.430&	0.065&	0.074   \\
  \\
  & & 0.4   &-0.773&	0.419&	0.359&	0.192&	0.096&	-0.039&	-0.149&	-0.488&	-0.578&	-0.727&	0.153&	0.257  \\
  \\
  &Rare & 1  &0.029 &  &  &  &  &  &  &  &  &  & &   \\
  \\
  & & 0.8   & -0.059&	0.468&	0.473&	0.341&	0.337&	0.055&	0.050&	-0.009&	-0.016 & & & \\
  \\
   & & 0.6  & -0.351&	0.798&	0.876&	0.558&	0.605&	0.171&	0.191&	-0.067&	-0.106 & & &  \\
  \\
   & & 0.4  & -0.674&	1.039&	1.379&	0.700&	0.816&	0.505&	0.263&	-0.448&	-0.857 & & &   \\
  \\
 $\Phi^{-1}(0.75)$ &Common & 1  & -0.011 &  &  &  &  &  &  &  &  &  & & \\
 \\
 & & 0.8  &-0.330&	0.045&	0.024&	-0.021&	-0.049&	-0.027&	-0.054&0.045&	0.025&-0.280&	-0.020&	-0.026 \\
  \\
  & & 0.6  &-0.623&	0.045&	-0.005&-0.117&	-0.187&	-0.162&	-0.225&	-0.167&	-0.216&	-0.568&	-0.152&	-0.170  \\
  \\
  & & 0.4  &-0.824&	0.098&	-0.004&	-0.145&	-0.271&	-0.297&	-0.430&-0.622&	-0.792&	-0.794&	-0.359&	-0.245   \\
  \\
  &Rare & 1  &  0.022 & &  &  & &  &  &  &  &  & &   \\
  \\
  & & 0.8  & -0.231&	0.351&	0.358&	0.163&	0.168&	0.014&	0.017&	-0.005&	-0.015 & & &  \\
  \\
   & & 0.6  & -0.505&	0.618&	0.682&	0.257&	0.302&	0.036&	0.073&	-0.016&	-0.044 & & &   \\
  \\
   & & 0.4  & -0.753&	0.804&	0.766&	0.315&	0.468&	0.195&	0.253&	-0.097&	-0.512  & & &  \\
  \\
 $\Phi^{-1}(0.9)$ &Common & 1  & -0.013 & &  &  & &  &  &  &  &  & &  \\
 \\
 & & 0.8  &  -0.393&-0.054&-0.066&	-0.090&-0.096&-0.098&-0.109&-0.104&	-0.112&	-0.322&	-0.066&	-0.065   \\
  \\
  & & 0.6  & -0.665&	-0.071&-0.109&	-0.173&-0.230&	-0.201&	-0.265&	-0.295&-0.394&	-0.600&	xxx&	-0.190  \\
  \\
  & & 0.4  &  -0.846&	-0.040&	-0.113&-0.194&-0.283&	-0.276&	-0.410&	-0.686&	-0.954&-0.814&	xxx&	-0.193  \\
  \\
  &Rare & 1  &  0.012 & &  &  & &  &  &  &  &  & &  \\
  \\
  & & 0.8  & -0.311&	0.216&	0.218&	0.075&	0.073&	0.010&	0.008&	0.002&	-0.014 &  & &  \\
  \\
   & & 0.6  &  -0.587&	0.372&	0.356&	0.111&	0.134&	0.024&	0.048&	-0.001&	-0.049 &  & &  \\
  \\
   & & 0.4  &  -0.804&	0.461&	0.104&	0.150&	*0.152&	0.072&	0.131&	-0.010&	-0.400&  & & \\
\end{tabular}
\end{center}
\footnotesize{$^a$ The values
in the cells are relative bias of the median, i.e., (median-0.405)/0.405. $^b$ (kn) indicates estimates under known nuisance parameters with $\sigma_w^2=1$ and with $\sigma_u^2$ that was determined according to the value of $\rho_{xw}$. $^c$ (ukn) indicates estimates under unknown nuisance parameters which were estimated by an external reliability sample of size 500 with 2 replications/pearson. $^d$ $n=3,000$ with cumulative incidence of 0.5. $^e$ $n=50,000$ with cumulative incidence of 0.03.$^f$ label of * denotes cases with convergence problems in Newton-Raphson, where $\left|\hat{\bth}\right|$ goes to infinity during the Newton-Raphson routine, and therefore the Newton-Raphson algorithm was limited to 100 iterations maximum.}
\end{landscape}


\begin{landscape}
\begin{center}
\fontsize{7.25}{0.25}\selectfont
\begin{tabular}{  m{1.1cm}  m{1.1cm} m{0.7cm} m{1.2cm} m{1.3cm} m{1.3cm} m{1.1cm} m{1.2cm} m{1.1cm} m{1.2cm} m{1.1cm} m{1.2cm} m{0.85cm} m{1.25cm} m{1.0cm}  }
\mc{10}{c}{Table 2: Finite Sample Bias$^a$ in $\omega$, ($\beta$, $\omega$)=(0.405, 0.693)} \\
$\tau$ &Disease& $\rho _{xw}$& Naive & RC1(kn)$^b$& RC1(ukn)$^c$ & RC2(kn)$^b$ & RC2(ukn)$^c$ & RR1(kn)$^b$ & RR1(ukn)$^c$ & RR2(kn)$^b$ &
RR2(ukn)$^c$& SIMEX &MPPLE(kn)$^b$&MPPLE(ukn)$^c$ \\
\\
$\Phi^{-1}(0.1)$ &Common & 1  & -0.342 &  & &  &  & &  &  &  &  & &  \\
 \\
 & & 0.8  & -1.044&	-1.069&	-0.954&-1.056&-0.936&	-0.903&	-0.846&-0.774&	-0.734&	-0.989&	-0.603&	\\
  \\
  & & 0.6  & -1.069&	-1.191&-1.160&-1.297&	-1.230&	-1.237&	-1.143&-0.703&	-0.636&	-1.039&	-0.829&	 \\
  \\
  & & 0.4  &-1.039&	-1.245&	-0.232&-1.361&	-1.286&	-1.253&-1.107&-0.561&	-0.555&	-1.018&	-0.880&	 \\
  \\
  &Rare & 1  &*-0.491$^d$ &  &  &  &  &  &  &  &  &  & &   \\
  \\
  & & 0.8  & -0.930	&-0.737&	-0.702&	-0.775&	-0.824&	-0.374&	-0.336&	0.458&	0.475 &  & &   \\
  \\
   & & 0.6  & -0.991&	0.008&	-0.288&	-0.923&	-0.836&	0.260&	0.525&	1.362&	1.486&  & &   \\
  \\
   & & 0.4  &-1.001&	1.741&	-0.672&	-0.709&	-0.952&	0.985&	0.240&	2.095&	0.684 &  & &   \\
  \\
 $\Phi^{-1}(0.25)$ &Common & 1  & 0.029 &  &  &  &  &  &  &  &  &  & &   \\
 \\
 & & 0.8  & -0.680&	-0.500&	-0.490&	-0.256&	-0.199&	-0.238&	-0.148&	-0.222&	-0.153&	-0.686&	-0.060&	-0.065 \\
   \\
  & & 0.6  & -0.908&-0.746&-0.692&	-0.597&-0.534&	-0.545&	-0.429&	-0.358&	-0.287&	-0.918&	-0.368&	-0.396  \\
  \\
  & & 0.4  & -0.974&	-0.840&-0.797&-0.811&	-0.714&	-0.715&-0.569&	-0.204&	-0.317&	-0.985&	-0.415&	-0.414 \\
  \\
  &Rare & 1  &-0.095 &  &  &  &  & &  &  &  &  & &   \\
  \\
  & & 0.8  & -0.707&	-0.413&	-0.400&	-0.384&	-0.367&	-0.244&	-0.243&	0.039&	0.040 &  & &  \\
  \\
   & & 0.6  & -0.919&	-0.685&	-0.612&	-0.592&	-0.551&	-0.353&	-0.214&	0.303&	0.299 &  & &  \\
  \\
   & & 0.4  & -0.984&	-0.836&	-0.770&	-0.752&	-0.649&	0.169&	0.146&	1.095&	0.599&  & & \\
  \\
 $\Phi^{-1}(0.5)$ &Common & 1  & 0.022 &  &  &  & &  & &  &  & & &   \\
 \\
 & & 0.8  & -0.562	&-0.317&	-0.328&	-0.077&-0.083&	-0.117&	-0.123&	-0.217&	-0.229&	-0.504&	0.090&	0.092   \\
  \\
  & & 0.6 & -0.820& -0.501& -0.476& -0.121& -0.093& -0.177& -0.150& -0.157& -0.168&  -0.815& 0.116& 0.103   \\
  \\
  & & 0.4   &-0.939&	-0.619&	-0.586&	-0.338&	-0.295&	-0.341&	-0.253&	0.022&	-0.072&	-0.947&	0.029&	-0.124   \\
  \\
  &Rare & 1  &-0.019 &  &  &  &  &  &  &  &  &  & &   \\
  \\
  & & 0.8   & -0.559&	-0.312&	-0.301&	-0.168&	-0.150&	-0.047&	-0.030&	-0.005&	-0.007 & & & \\
  \\
   & & 0.6  & -0.834&	-0.539&	-0.530&	-0.262&	-0.221&	-0.127&	-0.082&	0.025&	0.005 & & &  \\
  \\
   & & 0.4  & -0.955&	-0.720&	-0.660&	-0.325&	-0.217&	-0.326&	0.007&	0.260&	0.238 & & &   \\
  \\
 $\Phi^{-1}(0.75)$ &Common & 1  &0.032 &  &  &  &  &  &  &  &  &  & & \\
 \\
 & & 0.8  &-0.578&	-0.342&-0.347&	-0.080&	-0.099&	-0.136&	-0.151&	-0.195&	-0.226&	-0.519&	0.074&	0.072 \\
  \\
  & & 0.6  & -0.821&	-0.505&	-0.477&	-0.113&	-0.096&	-0.256&	-0.202&	-0.246&	-0.226&	-0.806&	0.060&	0.064  \\
  \\
  & & 0.4  & -0.937&	-0.604&	-0.537&	-0.322&-0.207&	-0.342&	-0.254&	-0.041&-0.131&	-0.941&	xxx&	-0.109  \\
  \\
  &Rare & 1  &  -0.057 & &  &  & &  &  &  &  &  & &   \\
  \\
  & & 0.8  & -0.503	&-0.232&	-0.215&	-0.005&	0.015&	-0.029&	-0.014&	-0.009&	-0.008 & & &  \\
  \\
   & & 0.6  & -0.790&	-0.400&	-0.361&	0.062&	0.144&	-0.060&	-0.020&	-0.006&	-0.033 & & &   \\
  \\
   & & 0.4  & -0.938&	-0.554&	-0.494&	0.133&	0.046&	-0.233&	-0.032&	0.067&	0.073  & & &  \\
  \\
 $\Phi^{-1}(0.9)$ &Common & 1  & 0.028 & &  &  & &  &  &  &  &  & &  \\
 \\
 & & 0.8  &  -0.673&	-0.489&	-0.484&	-0.231&	-0.241&	-0.323&	-0.331&	-0.273&	-0.293&	-0.668&	-0.177&	-0.174   \\
  \\
  & & 0.6  &-0.870&	-0.641&	-0.615&	-0.386&	-0.374&	-0.478&	-0.421&	-0.395&-0.315&	-0.876&	&	-0.311  \\
  \\
  & & 0.4  &  -0.955&	-0.716&	-0.647&	-0.538&	-0.452&	-0.558&	-0.463&	-0.214&	-0.160&	-0.967&&	-0.534 \\
  \\
  &Rare & 1  & -0.123 & &  &  & &  &  &  &  &  & &  \\
  \\
  & & 0.8  & -0.538&	-0.126&	-0.116&	0.099&	0.133&	-0.046&	-0.014&	-0.019&	-0.005 &  & &  \\
  \\
   & & 0.6  &  -0.817&	-0.190&	-0.249&	0.317&	0.494&	-0.104&	-0.022&	-0.011&	0.011 &  & &  \\
  \\
   & & 0.4  &  -0.949& 	0.531& 	-0.658& 	0.417& 	*0.065& 	-0.384& 	-0.069& 	0.123& 	0.244&  & & \\
\end{tabular}
\end{center}
\footnotesize{$^a$ The values
in the cells are relative bias of the median, i.e., (median-0.693)/0.693. $^b$ (kn) indicates estimates under known nuisance parameters with $\sigma_w^2=1$ and with $\sigma_u^2$ that was determined according to the value of $\rho_{xw}$. $^c$ (ukn) indicates estimates under unknown nuisance parameters which were estimated by an external reliability sample of size 500 with 2 replications/pearson. $^d$ $n=3,000$ with cumulative incidence of 0.5. $^e$ $n=50,000$ with cumulative incidence of 0.03.$^f$ label of * denotes cases with convergence problems in Newton-Raphson, where $\left|\hat{\bth}\right|$ goes to infinity during the Newton-Raphson routine, and therefore the Newton-Raphson algorithm was limited to 100 iterations maximum.}\\
\end{landscape}

\begin{landscape}
\begin{center}
\fontsize{7.25}{0.25}\selectfont
\begin{tabular}{  m{1.1cm}  m{1.1cm} m{0.7cm} m{1.2cm} m{1.3cm} m{1.3cm} m{1.1cm} m{1.2cm} m{1.1cm} m{1.2cm} m{1.1cm} m{1.2cm} m{0.85cm} m{1.35cm} m{0.9cm}  }
\mc{10}{c}{Table 3: Finite Sample Bias$^a$ in $\tau$, ($\beta$, $\omega$)=(0.405, 0.693)} \\
$\tau$ &Disease& $\rho _{xw}$& Naive & RC1(kn)$^b$& RC1(ukn)$^c$ & RC2(kn)$^b$ & RC2(ukn)$^c$ & RR1(kn)$^b$ & RR1(ukn)$^c$ & RR2(kn)$^b$ &
RR2(ukn)$^c$& SIMEX &MPPLE(kn)$^b$&MPPLE(ukn)$^c$ \\
\\
$\Phi^{-1}(0.1)$ &Common & 1  & -0.361 &  & &  &  & &  &  &  &  & &  \\
 \\
 & & 0.8  & -1.026&	-1.016&	-1.031& -1.053&	-1.041&	-1.040&	-1.016&	-1.071&	-1.022&	-1.040&	-0.829&	\\
  \\
  & & 0.6  & -1.080&	-1.029&	-1.057&-1.231&	-1.204&	-1.152&	-1.265&	-1.284&	-1.430&-1.043&	-0.931&	 \\
  \\
  & & 0.4  & -1.036&	-1.006&	-1.057&	-1.305&	-1.189&	-1.492&	-1.294&	-1.637&	-1.470&	-1.085&	-1.028&	\\
  \\
  &Rare & 1  &*-0.491$^d$ &  &  &  &  &  &  &  &  &  & &   \\
  \\
  & & 0.8  & -0.930	&-0.737&	-0.702&	-0.775&	-0.824&	-0.374&	-0.336&	0.458&	0.475 &  & &   \\
  \\
   & & 0.6  & -0.991&	0.008&	-0.288&	-0.923&	-0.836&	0.260&	0.525&	1.362&	1.486&  & &   \\
  \\
   & & 0.4  &-1.001&	1.741&	-0.672&	-0.709&	-0.952&	0.985&	0.240&	2.095&	0.684 &  & &   \\
  \\
 $\Phi^{-1}(0.25)$ &Common & 1  & -0.026 &  &  &  &  &  &  &  &  &  & &   \\
 \\
 & & 0.8  & -0.015&	-0.368&	-0.361&	0.176&	0.356&	-0.074&	0.059&	0.289&	0.420&	-0.572&	-0.162&	-0.151 \\
   \\
  & & 0.6  & -0.876&	-0.955&-0.875&	-1.051&	-0.749&	-0.911&	-0.528&	-0.777&	-0.212&	-0.821&	-0.754&	-0.853  \\
  \\
  & & 0.4  & -1.240&	-1.038&	-1.017&	-1.605&	-1.297&-1.936&-1.331&	-2.168&-1.544&	-0.952&	-1.077&	 \\
  \\
  &Rare & 1  &-0.095 &  &  &  &  & &  &  &  &  & &   \\
  \\
  & & 0.8  & -0.707&	-0.413&	-0.400&	-0.384&	-0.367&	-0.244&	-0.243&	0.039&	0.040 &  & &  \\
  \\
   & & 0.6  & -0.919&	-0.685&	-0.612&	-0.592&	-0.551&	-0.353&	-0.214&	0.303&	0.299 &  & &  \\
  \\
   & & 0.4  & -0.984&	-0.836&	-0.770&	-0.752&	-0.649&	0.169&	0.146&	1.095&	0.599&  & & \\
  \\
 $\Phi^{-1}(0.5)$ &Common & 1  & -0.006 &  &  &  & &  & &  &  & & &   \\
 \\
 & & 0.8  & -0.152	&-0.098&-0.080&	-0.187&-0.210&	0.056&	0.036&	0.020&	0.013&	-0.023&	0.039&	0.034   \\
  \\
  & & 0.6 &-0.126& -0.045&-0.071& -0.254& -0.316& 0.050& -0.010& 0.024& -0.053&  -0.080& 0.056& 0.070  \\
  \\
  & & 0.4   &0.109&	0.018&	0.005&	0.203&	0.094&	0.386&	0.192&	0.648&	0.296&	-0.031&	0.039&	0.074   \\
  \\
  &Rare & 1  &-0.019 &  &  &  &  &  &  &  &  &  & &   \\
  \\
  & & 0.8   & -0.559&	-0.312&	-0.301&	-0.168&	-0.150&	-0.047&	-0.030&	-0.005&	-0.007 & & & \\
  \\
   & & 0.6  & -0.834&	-0.539&	-0.530&	-0.262&	-0.221&	-0.127&	-0.082&	0.025&	0.005 & & &  \\
  \\
   & & 0.4  & -0.955&	-0.720&	-0.660&	-0.325&	-0.217&	-0.326&	0.007&	0.260&	0.238 & & &   \\
  \\
 $\Phi^{-1}(0.75)$ &Common & 1  & 0.007 &  &  &  &  &  &  &  &  &  & & \\
 \\
 & & 0.8  &-0.039&	-0.384&	-0.410&	-0.079&	-0.158&	0.256&	0.146&	0.363&	0.244&	-0.310&	0.035&	0.021 \\
  \\
  & & 0.6  & -0.273&	-0.738&	-0.846&	-0.294&	-0.587&	0.075&	-0.181&0.470&	0.138&	-0.603&	-0.360&	-0.386  \\
  \\
  & & 0.4  & -1.001&	-1.000&	-0.970&	-0.347&	-0.717&	-0.193&	-0.630&	0.016&	&	-0.964&	&	-0.834  \\
  \\
  &Rare & 1  &  -0.057 & &  &  & &  &  &  &  &  & &   \\
  \\
  & & 0.8  & -0.503	&-0.232&	-0.215&	-0.005&	0.015&	-0.029&	-0.014&	-0.009&	-0.008 & & &  \\
  \\
   & & 0.6  & -0.790&	-0.400&	-0.361&	0.062&	0.144&	-0.060&	-0.020&	-0.006&	-0.033 & & &   \\
  \\
   & & 0.4  & -0.938&	-0.554&	-0.494&	0.133&	0.046&	-0.233&	-0.032&	0.067&	0.073  & & &  \\
  \\
 $\Phi^{-1}(0.9)$ &Common & 1  & -0.006 & &  &  & &  &  &  &  &  & &  \\
 \\
 & & 0.8  &  -0.223&	-0.502&	-0.524&	-0.094&	-0.110&	-0.010&	-0.055&	0.144&	&	-0.539&-0.187&	-0.187   \\
  \\
  & & 0.6  & -0.593&	-0.853&	-0.911&	-0.529&	-0.651&	-0.407&-0.571&	-0.209&	&	-0.850&	&	-0.661 \\
  \\
  & & 0.4  &  -1.048&	-1.008&	-0.989&	-0.877&	-0.808&	-0.796&	-0.786&	-0.647&	&	-0.904&	&	-0.932  \\
  \\
  &Rare & 1  & -0.123 & &  &  & &  &  &  &  &  & &  \\
  \\
  & & 0.8  & -0.538&	-0.126&	-0.116&	0.099&	0.133&	-0.046&	-0.014&	-0.019&	-0.005 &  & &  \\
  \\
   & & 0.6  &  -0.817&	-0.190&	-0.249&	0.317&	0.494&	-0.104&	-0.022&	-0.011&	0.011 &  & &  \\
  \\
   & & 0.4  &  -0.949& 	0.531& 	-0.658& 	0.417& 	*0.065& 	-0.384& 	-0.069& 	0.123& 	0.244&  & & \\
\end{tabular}
\end{center}
\footnotesize{$^a$ The values
in the cells are relative bias of the median, i.e., (median-$\tau$)/$\tau$. $^b$ (kn) indicates estimates under known nuisance parameters with $\sigma_w^2=1$ and with $\sigma_u^2$ that was determined according to the value of $\rho_{xw}$. $^c$ (ukn) indicates estimates under unknown nuisance parameters which were estimated by an external reliability sample of size 500 with 2 replications/pearson. $^d$ $n=3,000$ with cumulative incidence of 0.5. $^e$ $n=50,000$ with cumulative incidence of 0.03.$^f$ label of * denotes cases with convergence problems in Newton-Raphson, where $\left|\hat{\bth}\right|$ goes to infinity during the Newton-Raphson routine, and therefore the Newton-Raphson algorithm was limited to 100 iterations maximum.}\\
\end{landscape}
\subsection{Background}
\noindent When the changepoint $\tau$ is unknown, we need to estimate not only $\beta$ and $\omega$ but $\tau$ as well. Since there are discontinuities in the derivative of the log partial likelihood with respect to $\tau $, conventional maximization techniques cannot be applied. The obvious approach is to do a grid search over $\tau $. A more efficient possible approach would be to do a bisection search over $\tau$; this approach would work if the log-likelihood could be assumed to be monotone on each side of the maximum point. We examined the plausibility of this assumption under the changepoint Cox model without measurement error, by simulating data according to this model and plotting the log-likelihood as a function of $\tau$ for various values of $\beta $ and $\omega $, including the true values used to generate the data and a few other values. Examining the graphs, it seems that in most cases there is a common up-and-down shape in the log-likelihood. There are some cases, with $\beta =5$, where the function has a blip. Further investigation revealed that for very high values of $\beta $, Matlab runs into an overflow problem. We therefore eliminated these cases and restricted attention to cases where $\beta X$ and $\omega \, {\rm (}X{\rm -}\tau )_{+} $ lie in a reasonable range of [-20,20]. When we imposed this restriction, we no longer observed any blips, and the log-likelihood functions took two possible forms (depending on the specific parameter values): 1. Completely monotone, always ascending or descending, with the optimum at the end of the range. 2. Up-and-down pattern with a single maximum at the center.

\noindent It is thus justified to use a bisection search rather than a grid search to find the estimate of $\tau $. For any given value of $\tau $, we estimated $\beta $ and $\omega $ using the methods described previously. We then carried out a bisection search to find the value of $\tau $ that maximizes the relevant objective function. The bisection search over $\tau $ was carried out in Matlab using the function \textit{fminbnd}, and was done over a specified range. For the na\"{i}ve method, the range spanned from the \textit{q}-th to (1-\textit{q})-th quantile of the observed $W_{i} $ values, where we took q=0.05. For RC1, RC2, and RR, the range spanned from the \textit{q}-th to the (1-\textit{q})-th quantile of the sample values of $E[X_{i} |W_{i} ]$, again taking \textit{q}=0.05.

\indent In the RR method with unknown nuisance parameters (RR(unk)), there were some problematic cases for which the estimated $\sigma _{w}^{2} $ was negative and we put zero value instead. Then the values of $E[X_{i} |W_{i} ]$ (as a function of the estimated $\sigma _{w}^{2} $) become the same value for each subject and therefore equal values of $\tau _{1} ,\tau _{2} $ as percentiles of $E[X_{i} |W_{i} ]$ and problem in the bisection method. Therefore for these problematic problems, which are labeled with **, we took $\tau _{1} ,\tau _{2}$ as percentiles of $W$. (Note that in comparing the results of RR by taking $\tau _{1} ,\tau _{2} $ as percentiles of $E[X_{i} |W_{i} ]$ with the results of RR by taking $\tau _{1} ,\tau _{2} $ as percentiles of $W$, the results are better when considering percentiles of $E[X_{i} |W_{i} ]$).
\noindent The starting values of $\beta $ and $\omega $ were selected as in Agami et al. (2018).

\indent In the SIMEX method we took additional datasets $B=200$. We examined the extrapolation functions of  first and second degree polynomial. The performance of the first-degree polynomial extrapolant was better and the reports results of the SIMEX are based on this extrapolant.

\subsection{Summary of Simulation Results} 
\noindent A number of common trends were seen for all the correction methods examined. The estimator of $\beta$ was under-estimated under all methods when $\tau >0$. This bias became lower as $\tau $ increased for all methods except the naive and the RC1 methods.

\noindent The estimator of $\omega $ was over-estimated when $\tau >0$ under all the methods except the naive and the RC1 methods. This bias became lower at most cases as the error variance increased.

\noindent The estimator of $\tau $ performed well when $\tau $was in the middle of the covariate domain ($\tau $=0), and substantially less well when the changepoint was at the upper or lower extreme of the covariate domain. The estimator of $\tau $ performed well in the rare disease case than in the common disease case.
\noindent As expected, the estimators performed progressively less well as the error variance increased. The standard deviation of the estimators decreased as $\tau $ increased.

\indent The best performing method in the common disease was the MPPLE, and the best performing method in the rare disease case was RR2. The RC2 method was better than the MPPLE in $\tau =-1.28$ in the estimators of $\beta $ and $\omega $. The RC2 estimator of $\beta$ performed better than the RC1 for $\tau <0$ and worse for $\tau \ge 0$ for all values of the error variance, under known and unknown nuisance parameters, and under common and rare disease. The RC2 estimator of $\omega $ was better than the RC1 estimator in most cases. The RC2 estimator of $\omega $ in the common disease case was better than the naive estimator, but the contrary for $\beta $ in $\tau \le -0.67$. The RC2 estimators of $\omega $ in the rare disease case was better than the naive estimator in $\tau \le -0.67$ and the contrary for $\beta $. The naive estimator of $\tau $ was better than the RC2 estimator in most cases. The SIMEX estimator under common disease was slightly better than the naive method for the estimator of $\beta $ for $\tau \ge 0$, but the naive estimator was better than the SIMEX estimator for the estimators of $\omega $ and $\tau $.


\subsection{Robustness} 
One of the assumptions we had (Chapter 2.1) is the normality assumption of the additive measurement error model. We checked the robustness of this
assumption: We rerun the simulations with \emph{t}-distribution instead of the normal distribution, with degrees of freedom (df) of 6 and 15. We matched the mean and the variance to the that used for the normal distribution in the simulations. The results of  \emph{t}-distribution with df=15 were closed to the results that were based on the normal distribution, whereas the results of \emph{t}-distribution with df=6 were far from the results that were based on the normal distribution. That is, the normality assumption is important in the applications of the methods we considered.
\section{Illustrative Example}
As noted in the introduction, our work was motivated by some instances of threshold effects observed in the Nurses' Health Study (NHS), including threshold effects observed in the NHS's investigation of the long term health effects of air pollution. We consider an analysis of the NHS concerning the effect of exposure to particulate matter of diameter 10 $\mu$m or less (PM$_{10}$) in relation to fatal myocardial infarction (MI). Here, 93,013 female nurses were followed from June 1992 to June 2006, with 1,073 fatal MI events observed. PM$_{10}$ exposure was assessed for each individual by linking the individual's residential address to her predicted PM$_{10}$ exposure using a spatio-temporal model derived from data from EPA area monitors (Yanosky et al.,
2008; Paciorek et al., 2009). The time scale in the analysis was age in months, so that the data are subject to left-truncation.

We fit a stratified Cox model, with strata defined by age in months. Aside from the main covariate PM$_{10}$ , the model included calendar year (yr), indicator variables for season, and indicator variables for US state of residence, all time-varying.
We applied all the methods we discussed above except the SIMEX , RR2 and MPPLE methods, which have a heavy computational burden. We took $\tau_{min}$ and $\tau_{max}$ to be equal,respectively, to the 15th and 75th percentiles of the observed distribution of PM$_{10}$ across the entire person-time experience, and we assumed that the threshold were known in advance. To estimate the conditional expectation $E(X|W)$ and conditional variance $Var(X|W)$ need for the correction methods, we used an external validation study of 98 person-months in 4 cities of personal PM$_{10}$ measurements, taken using personal environmental monitors and a surrogate exposure based on the spatio-temporal model of Yanosky (Kioumourtzoglou et al.(2014)). We fit a mixed linear model of personal PM$_{10}$ on the surrogate exposure, and we obtained $E(X|W)=0.9737+ 0.6349W $ and $Var(X|W)=0.5817$. Table 4 summarizes the results of the analyses. We report the results for PM$_{10}$ and for (PM$_{10} -\tau)_+$ (which labeled PM$_{10}{\tau}$) only, but we corrected for the measurement error in the additional covariates of season, year and and the state covariates, as well. We giving the estimate, the p-value in bracket, and the standard error. The estimate of $\omega$ was significant under most of the methods considered including the RR for the 10th percentile of PM$_{10}$, whereas it was significant for the 10th percentile of PM$_{10}$ under all methods except the RR. That is, there is a possible changepoint at one of these percentiles. The estimates of the additional covariates of season, year and state were significant at a significant level of 0.05.
\begin{landscape}
\begin{center}
\fontsize{8.5}{0.5}\selectfont
\
\begin{tabular}{llcccccccccccccccc}
\mc{13}{c}{Table 4: Illustrative Example} \\
\mc{15}{c}{Results for the NHS Study of Chronic PM$_{10}$ Exposure in Relation to Fatal MI} \\
\mc{15}{c}{Cox Analyses Assuming Known Threshold $\tau$$^a$} \\
\\
\\
$\tau^*$ (\%ile)$^b$ &  covariate & Naive & RC1 & RC2 & RR  \\
\\
10th &PM$_{10}$  &1.177(0.389)  &6.427(42.897) &2.734(1.346)       &3.045(1.390) \\
                    &&0.003        &0.881      &0.042             &0.029 \\
                    &&[0.414,1.940]&[-77.651,90.505]&[0.096,5.372] &[0.321,5.769]\\
                    &&[1.513,6.959]& [0.000,2E+39]&[1.101,215.293] &[1.379,320.217]\\
\\
\\

              &PM$_{10}$${\tau}$$^c$&-1.021(0.393) &-6.121(42.988)    &-2.733(1.462)    &-2.875(1.421)\\
                    &&0.009            &0.887                          &     0.062                   &0.043 \\
                    &&[-1.791,-0.251]& [-90.377,78.135] &[-5.599,0.133] &[-5.660,-0.090]\\
                    &&[0.167,0.778]  &[0.000,9E+33]     &[0.004,1.142] &[0.003,0.914]\\
\\
\\
25th &PM$_{10}$ &0.788(0.162)   &4.846(3.295)      &1.928(0.858)    &2.902(1.452)\\
     &&         (0.000)         &0.141             &0.025           &0.046\\
     &&        [0.470,1.106]    &[-1.612,11.304]   &[0.246,3.610]   &[0.056,5.748]\\
     &&        [1.600,3.022]    &[0.199, 81145.57] &[1.279,36.966]  &[1.058,313.563]\\
     \\
     \\
     &PM$_{10}$${\tau}$ &-0.664(0.180)  &-4.566(0.173)    &-1.951(1.005)    &-2.811(1.500)\\
     &&                 0.000           &3.351            &0.052            &0.061\\
     &&                 [-1.017,-0.311] &[-11.134,2.002]  &[-3.921,0.019]   &[-5.751,0.129]\\
     &&                 [0.362,0.733]   &[0.000,7.404]    &[0.020,1.019]    &[0.003,1.138]\\
     \\
     \\
50th &PM$_{10}$  &0.518(0.154)  &1.394(0.423)     &1.325(0.326)   &2.165(1.408)\\
                &&0.001         &0.001            &0.000          &0.124  \\
                &&[0.216,0.820] &[0.565,2.223]    &[0.686,1.964]  &[-0.595,4.925]\\
                &&[1.241,2.270] &[1.759,9.235]    &[1.986,7.128]  &[0.552,137.689]\\
\\
\\
              &PM$_{10}$${\tau}$&-0.435(0.191) &-1.180(0.458)    &-1.383(0.437)   &-2.181(1.507)\\
                    &&          0.022          &0.010            &0.002           &0.148 \\
                    &&          [-0.809,-0.061]&[-2.078,-0.282]  &[-2.240,-0.526] &[-5.135,0.773]\\
                    &&          [0.445,0.941]  &[0.125,0.754]    &[0.106,0.591]   &[0.006,2.166] \\
\\
\\
75th &PM$_{10}$ &0.353(0.100)  &0.635(0.188) &0.915(0.008)    &1.298(0.863)\\
               &&0.000         &0.001        &0.346           &0.132\\
               &&[0.157,0.549] &[0.267,1.003]&[0.899,0.931]   &[-0.393,2.989]\\
               &&[1.170,1.732] &[1.306,2.726]&[2.457,2.537]   &[0.675,19.866]\\
\\
\\
&PM$_{10}$${\tau}$ &-0.299(0.154)    &-0.528(0.259)    &-1.025(0.562)    &-1.407(1.068)\\
                  &&0.052            &0.042            &0.068            &0.188\\
                  &&[-0.601,0.003]   &[-1.036,-0.020]  &[-2.127,0.077]   &[-3.500,0.686]\\
                  &&[0.548,1.003]    &[0.355,0.980]    &[0.119,1.080]  &[0.030,1.986]\\

\end{tabular}
\end{center}
\footnotesize{$^a$ each cell contains (in that order): estimate(standard deviation), p-value, $95\%$ confidence interval of $\exp$(relevant coefficient), $95\%$ confidence interval of $\exp$($10\times$(relevant coefficient)).
$^b$ percentile of PM$_{10}$. $^c$ PM$_{10}$${\tau}$ = (PM$_{10} -\tau)_+$.}\\
\end{landscape}


\section{Summary}

We have developed point and interval estimators of the regression coefficients in a Cox survival model with a changepoint, in a
setting where the covariate values are subject to measurement error. This type of analysis is of interest in many
epidemiological studies. We considered the case where the changepoint is known and where the covariate of main interest is of one-dimension. All the methods developed in this paper can be extended to a multi-dimensional case. In addition to the naive method
ignoring the measurement error, we examined the following methods: regression calibration (in two versions, RC1 and RC2), SIMEX,
the induced
relative risk approach of Prentice (1982) (in two versions: Prentice's original proposal (RR1) and a version with a bootstrap
bias correction (RR2)), and the MPPLE method of Zucker (2005). 

We derived the asymptotic properties of the estimators, and carried
out a simulation study under rare and common disease settings to compare them with respect to bias and confidence
interval coverage. The simulation study considered the rare and common disease settings and range of values for the correlation
between the true covariate value and the measured value. In general, all the methods performed better than the naive method, but
the best performing methods were the RR2 and the MPPLE methods.

The deviation of the estimator of $\omega$ from its true value of 0.69 for $\rho_{xw}$=0.8 was between -0.08 to 0.02 in RR2 and MPPLE methods, compared with -0.51 to 0.03 in the naive, SIMEX and the RC methods, in the common disease. This deviation was between 0
to 0.10 in RR2 method and between -0.52 to 0.07 in the naive and RC methods, in the rare disease.
The key factors determining the performance of the methods was the correlation between the true variable and its surrogate and
the location of the changepoint. As expected, the estimators performed better with less measurement error and a centrally located
changepoint.

\newpage



\section{Proof of Results}
\noindent \textbf{Proof of Lemma 1.} Since the score is continuously differentiable in the first three components, we have that $\sum _{i=1}^{n}\bU_{j} \left(\bV_{i} ,\hat{\bth },g\right) =0$ for $j=1,\, 2,\, 3$, where
\[\bU_{i} \left(\bV_{i} ,\bth ,g\right)=\delta _{i} \left[\bxi _{i} \left(T_{i}^{0} ,\bth ,g\right)-\frac{S^{\left(1\right)} \left(T_{i}^{0} ,\bth ,g\right)}{S^{\left(0\right)} \left(T_{i}^{0} ,\bth ,g\right)} \right].\]
In regard to the fourth component:

\noindent Let $F\left(a\right)=\sum _{i=1}^{n}F_{i} \left(a\right) $ with
\[F_{i} \left(a\right)=\delta _{i} \left[\begin{array}{l} {\left(-\omega \right)\cdot {\rm I} \left\{g_{1} \left(W_{i} \left(T_{i}^{0} \right)\right)\ge \hat{\tau }+a\right\}} \\ {-\frac{\sum _{j=1}^{n}Y_{j} \left(T_{i}^{0} \right)e\left(W_{j} \left(T_{i}^{0} \right),\bZ_{j} \left(T_{i}^{0} \right),\, \left(\hat{\bgam }^{T} ,\hat{\beta },\hat{\omega },\hat{\tau }+a\right)\right)\left(-\omega \right)\cdot {\rm I} \left\{g_{1} \left(W_{j} \left(T_{i}^{0} \right)\right)\ge \hat{\tau }+a\right\} }{\sum _{j=1}^{n}Y_{j} \left(T_{i}^{0} \right)e\left(W_{j} \left(T_{i}^{0} \right),\bZ_{j} \left(T_{i}^{0} \right),\, \, \left(\hat{\bgam }^{T} ,\hat{\beta },\hat{\omega },\hat{\tau }+a\right)\right) } } \end{array}\right]\, .\]
The function $F\left(a\right)$ is left continuous with jumps at the points $a=g_{1} \left(W_{i} \left(T_{i}^{0} \right)\right)-\hat{\tau }$.

\noindent Since $\hat{\tau }$ is a maximum point of $l_{p} \left(\bth ,g\right)$ , we have that $F\left(0^{-} \right)=\mathop{\lim }\limits_{a\uparrow 0} F\left(a\right)\ge 0$ and $F\left(0^{+} \right)=\mathop{\lim }\limits_{a\downarrow 0} F\left(a\right)\le 0$. Thus,
\[\begin{array}{l} {\left|F\left(0\right)\right|=\left|F\left(0^{-} \right)\right|\le \left|F\left(0^{-} \right)-F\left(0^{+} \right)\right|=\sum _{i=1}^{n}\left(F_{i} \left(0^{-} \right)-F_{i} \left(0^{+} \right)\right) } \\ {=-\sum _{i=1}^{n}\delta _{i}  \left[\left(-\omega \right)-\frac{\sum _{j=1}^{n}Y_{j} \left(T_{i}^{0} \right)r\left(g_{1} \left(W_{j} \left(T_{i}^{0} \right)\right),Z_{j} \left(T_{i}^{0} \right),\, \left(\hat{\bgam }^{T} ,\hat{\beta },\hat{\omega },\hat{\tau }\right)\right)\left(-\omega \right) }{\sum _{j=1}^{n}Y_{j} \left(T_{i}^{0} \right)e\left(g_{1} \left(W_{j} \left(T_{i}^{0} \right)\right),Z_{j} \left(T_{i}^{0} \right),\, \left(\hat{\bgam }^{T} ,\hat{\beta },\hat{\omega },\hat{\tau }\right)\right) } \right]{\rm I} \left\{g_{1} \left(W_{i} \left(T_{i}^{0} \right)\right)=\hat{\tau }\right\}\, .} \end{array}\]
(that is, the difference of $F_{i} \left(0^{-} \right)-F_{i} \left(0^{+} \right)$ equals zero unless $g_{1} \left(W_{i} \left(T_{i}^{0} \right)\right)=\hat{\tau }$).

\noindent By assumptions GA1 and GA2, the expression  $\left[\left(-\omega \right)-\frac{\sum _{j=1}^{n}Y_{j} \left(T_{i}^{0} \right)e\left(g_{1} \left(W_{j} \left(T_{i}^{0} \right)\right),\bZ_{j} \left(T_{i}^{0} \right),\, \left(\hat{\bgam }^{T} ,\hat{\beta },\hat{\omega },\hat{\tau }\right)\right)\left(-\omega \right) }{\sum _{j=1}^{n}Y_{j} \left(T_{i}^{0} \right)r\left(g_{1} \left(W_{j} \left(T_{i}^{0} \right)\right),\bZ_{j} \left(T_{i}^{0} \right),\, \left(\hat{\bgam }^{T} ,\hat{\beta },\hat{\omega },\hat{\tau }\right)\right) } \right]$ is bounded by some constant $K$, and since $\delta _{i} $ is bounded by 1, we have
 \\
 $\begin{array}{l} {-\sum _{i=1}^{n}\delta _{i}  \left[\left(-\omega \right)-\frac{\sum _{j=1}^{n}Y_{j} \left(T_{i}^{0} \right)e\left(g_{1} \left(W_{j} \left(T_{i}^{0} \right)\right),\bZ_{j} \left(T_{i}^{0} \right),\, \left(\hat{\bgam }^{T} ,\hat{\beta },\hat{\omega },\hat{\tau }\right)\right)\left(-\omega \right) }{\sum _{j=1}^{n}Y_{j} \left(T_{i}^{0} \right)r\left(g_{1} \left(W_{j} \left(T_{i}^{0} \right)\right),\bZ_{j} \left(T_{i}^{0} \right),\, \left(\hat{\bgam }^{T} ,\hat{\beta },\hat{\omega },\hat{\tau }\right)\right) } \right]{\rm I} \left\{g_{1} \left(W_{i} \left(T_{i}^{0} \right)\right)=\hat{\tau }\right\}} \\ {\le K\sum _{i=1}^{n}{\rm I} \left\{g_{1} \left(W_{i} \left(T_{i}^{0} \right)\right)=\hat{\tau }\right\} .} \end{array}$

\noindent By assumption GA2, $W_{i} \left(T_{i}^{0} \right)$ has a continuous distribution, and therefore
\\
$P\left(g_{1} \left(W_{i} \left(T_{i}^{0} \right),g_{1} \left(W_{j} \left(T_{i}^{0} \right)\right)\right)=\hat{\tau },\, for\, i\ne j\right)=0$, that is, the indicator term in the sum  $\sum _{i=1}^{n}{\rm I} \left\{g_{1} \left(W_{i} \left(T_{i}^{0} \right)\right)=\hat{\tau }\right\} $ equals 1 for one subject only.

\noindent Therefore, $n^{-{1 \mathord{\left/{\vphantom{1 2}}\right.\kern-\nulldelimiterspace} 2} } \left|F\left(0\right)\right|\le n^{-{1 \mathord{\left/{\vphantom{1 2}}\right.\kern-\nulldelimiterspace} 2} } K\mathop{\to }\limits_{n\to \infty } 0\, .$ Then, similar to the argument in the thesis of K\"{u}chenhoff (1995) page 28, we conclude that $F\left(0\right)=\sum _{i=1}^{n}\bU_{4} \left(V_{i} ,\hat{\bth },g\right) \mathop{\to }\limits^{p} 0$. $\square$

\noindent \textbf{Proof of Lemma 2. }

\noindent  (N-2) (i) We can write $\tilde{Q}\left(\bth \right)=\tilde{Q}\left(\bth ^{*} \right)+\bJ\left(\tilde{\bth }\right)\left(\bth -\bth ^{*} \right)=\bJ\left(\tilde{\bth }\right)\left(\bth -\bth ^{*} \right)$, where $\bJ$ is the Jacobian matrix of $\tilde{Q}\left(\bth \right)$, and $\tilde{\bth }$ is between $\bth $ and $\bth ^{*} $.
Let $\eta _{\min } \left(\bth \right)$ be the minimal eigenvalue of $\bJ\left(\bth \right)^{T} \bJ\left(\bth \right)$.
By assumption, $\bJ$ is nonsingular  at $\bth ^{*} $ and hence $\eta _{\min } \left(\bth ^{*} \right)>0$.
Since $J\left(\bth \right)$ is continuous and the eigenvalues of a matrix are continuous functions of the elements of the
matrix, for $d_{0}$ sufficiently small we have $\bar{\eta }_{\min }
=\mathop{\min }\limits_{\left\| \bthsub -\bthsub ^{*} \right\| \le d_{0} } \eta _{\min } \left(\bth \right)>0$.
We thus get
$$
\left\| \tilde{Q}\left(\bth \right)\right\| ^{2} =\left(\bth -\bth ^{*} \right)^{T} J\left(\tilde{\bth }\right)^{T} \bJ\left(\tilde{\bth }\right)\left(\bth -\bth ^{*} \right)\ge \bar{\eta }_{\min } \left\| \bth -\bth ^{*} \right\| ^{2}
$$
\noindent (ii) Let $\bth _{1} ,\bth _{2} $ such that $\left\| \bth _{1} -\bth _{2} \right\| \le d$. We have
\[\begin{array}{l} {\left\| \tilde{\bvarphi}\left(v,\bth _{1} ,g\right)-\tilde{\bvarphi}\left(v,\bth _{2} ,g\right)\right\|
=\left\| \bxi \left(T^{0} ,\bth _{1} ,g\right)
-\frac{s^{\left(1\right)} \left(T_{}^{0} ,\bth _{1} ,g\right)}{s^{\left(0\right)} \left(T_{}^{0} ,\bth _{1} ,g\right)}
-\bxi \left(T^{0} ,\bth _{2} ,g\right)
+\frac{s^{\left(1\right)} \left(T_{}^{0} ,\bth _{2} ,g\right)}{s^{\left(0\right)} \left(T_{}^{0} ,\bth _{2} ,g\right)} \right\| } \\ {\le \left[\left\| \bxi \left(T^{0} ,\bth _{1} ,g\right)-\bxi \left(T^{0} ,\bth _{2} ,g\right)\right\| +\left\| \frac{s^{\left(1\right)} \left(T_{}^{0} ,\bth _{2} ,g\right)}{s^{\left(0\right)} \left(T_{}^{0} ,\bth _{2} ,g\right)} -\frac{s^{\left(1\right)} \left(T_{}^{0} ,\bth _{1} ,g\right)}{s^{\left(0\right)} \left(T_{}^{0} ,\bth _{1} ,g\right)} \right\| \right]\equiv
\left[\left\| \left(1\right)\right\| +\left\| \left(2\right)\right\| \right].} \end{array}\]
Regarding (1), the only component of $\bxi$ requiring attention is the last one. We have
$$
\begin{array}{l}
{\left\| \xi_{p+2} \left(T^{0} ,\bth _{1} ,g\right)-\xi_{p+2} \left(T^{0} ,\bth _{2} ,g\right)\right\| } \\ {=\left\| \left(g_{1} \left(W\right)-\tau _{1} \right)_{+} -\left(g_{1} \left(W\right)-\tau _{2} \right)_{+} ,\, \left(-\omega _{1} \right){\rm I} \left(g_{1} \left(W\right)>\tau _{1} \right)-\left(-\omega _{2} \right){\rm I} \left(g_{1} \left(W\right)>\tau _{2} \right)\right\| \, .}
\end{array}
$$
Recall the definition $g_2(w,\tau) = (g_1(w)-\tau)_+$.  For $\, \tau _{1} <\tau _{2} $ we have
$$
\begin{array}{l} {\left[g_{2} \left(\tau _{1} \right)-g_{2} \left(\tau _{2} \right)\right]
=\left\{\begin{array}{cc} {0} & {g_{1} <\tau _{1} } \\
{\left(g_{1} -\tau _{1} \right)} & {g_{1} \in \left[\tau _{1} ,\tau _{2} \right]} \\
{\left[\left(g_{1} -\tau _{1} \right)-\left(g_{1} -\tau _{2} \right)\right]
=\left(\tau _{2} -\tau _{1} \right)} & {g_{1} >\tau _{2} } \end{array}\right. \,}
\end{array}
$$

Hence
$|g_2(\tau_2) - g_2(\tau_1)| \leq |\tau_2 - \tau_1|$.
By symmetry this holds for $\tau _{1} >\tau _{2} $ as well.
Therefore,
$\left\| \left(g_{1} \left(W\right)-\tau _{1} \right)_{+} -\left(g_{1} \left(W\right)-\tau _{2} \right)_{+} \right\| \le \left|\tau _{2} -\tau _{1} \right|\le \left\| \bth _{2} -\bth _{1} \right\| .$
In regard to the second difference: If $\tau _{1} <\tau _{2} $, then
\[\left(-\omega _{1} \right){\rm I} \left(g_{1} \left(W\right)>\tau _{1} \right)-\left(-\omega _{2} \right){\rm I} \left(g_{1} \left(W\right)>\tau _{2} \right)=\left\{\begin{array}{cc} {0} & {g_{1} \left(W\right)<\tau _{1} } \\ {-\omega _{1} } & {g_{1} \left(W\right)\in \left[\tau _{1} ,\tau _{2} \right]} \\ {-\omega _{1} +\omega _{2} } & {g_{1} \left(W\right)>\tau _{2} } \end{array}\right.  .\]
Thus,
\begin{align*}
& {E\left[\sup \left|\left(-\omega _{1} \right){\rm I} \left(g_{1} \left(W\right)>\tau _{1} \right)-\left(-\omega _{2} \right)
{\rm I} \left(g_{1} \left(W\right)>\tau _{2} \right)\right|\right]} \\
& \hspace*{36pt} = {\omega _{1} P\left(g_{1} \left(W\right)\in \left[\tau _{1} ,\tau _{2} \right]\right)+\left(\omega _{2} -\omega _{1} \right)P\left(g_{1} \left(W\right)>\tau _{2} \right)\, .}
\end{align*}
Now, $P\left(W\in \left[\tau _{1} ,\tau _{2} \right]\right)\le \tilde{M}_{W} \left(\tau _{2} -\tau _{1} \right)$, where $\tilde{M}_{W}$ is the maximum of the density of  $W$.
It follows that $P\left(g_{1} \left(W\right)\in \left[\tau _{1} ,\tau _{2} \right]\right)\le M_{W} \left(\tau _{2} -\tau _{1} \right)$, where $M_{W} $ is some constant.
Therefore,
$$
E\left[\sup \left|\left(-\omega _{1} \right){\rm I} \left(g_{1} \left(W\right)>\tau _{1} \right)-\left(-\omega _{2} \right)
{\rm I} \left(g_{1} \left(W\right)>\tau _{2} \right)\right|\right]
\le M_{W}^{\prime \prime} \left\| \bth _{2} -\bth _{1} \right\|
$$
where $M_{W}^{\prime \prime} $ is some constant. By symmetry this is also true for $\tau _{1} >\tau _{2} $.
Therefore,
${E\left(\left\| \left(1\right)\right\| \right)}$ ${\le M_{W}^{''} \left\| \theta _{2} -\theta _{1} \right\|}$.

\noindent Term $\left(2\right)$: It suffices to show that
${s^{\left(1\right)} \left(T_{}^{0} ,\bth ,g\right)}/{s^{\left(0\right)} \left(T_{}^{0} ,\bth,g\right)}$
has a bounded derivative. This follows from the fact that
$s^{\left(j\right)} \left(t,\bth ,g\right),\, j=0,1$ is differentiable and its first
derivative is bounded (because the covariates and parameters are bounded by Assumptions GA1 and GA2)
and the fact that ${s^{\left(0\right)} \left(T_{}^{0} ,\bth,g\right)}$ is bounded below.

\noindent  (iii) Same as the proof of (ii).

\noindent (N-4) This follows from the boundedness assumption on $\bV\left(g\left(t,\tau \right)\right)$. $\square$

\noindent \textbf{Proof of Theorem} \textbf{1.}

\noindent We can write
\[\begin{array}{l} {\tilde{\bPs }_{n} \left(\bth ^{*} \right)+\bPs _{n} \left(\bth ^{*} \right)-\tilde{\bPs }_{n} \left(\bth ^{*} \right)+\tilde{\bQ}\left(\hat{\bth }_{n} \right)}  =\\{-[\tilde{\bPs }_{n} \left(\hat{\bth }_{n} \right)-\tilde{\bPs }_{n} \left(\bth ^{*} \right)-\tilde{\bQ}\left(\hat{\bth }_{n} \right)]+\bPs _{n} \left(\hat{\bth }_{n} \right)-\left[\left(\bPs _{n} \left(\hat{\bth }_{n} \right)-\tilde{\bPs }_{n} \left(\hat{\bth }_{n} \right)\right)-\left(\bPs _{n} \left(\bth ^{*} \right)-\tilde{\bPs }_{n} \left(\bth ^{*} \right)\right)\right]}\\ {+\bPs _{n} \left(\hat{\bth }_{n} \right).} \end{array}\]
\noindent Define
\begin{equation}
\tilde{\bZ}_{n} \left(\zeta ,\bth \right)=\frac{\left(\tilde{\bPs }_{n} \left(\zeta \right)-\tilde{\bPs }_{n} \left(\bth \right)\right)-\left(\tilde{\bQ}\left(\zeta \right)-\tilde{\bQ}\left(\bth \right)\right)}{n^{{-1 \mathord{\left/{\vphantom{-1 2}}\right.\kern-\nulldelimiterspace} 2} } +\left\| \tilde{\bQ}\left(\zeta \right)\right\| }.
\label{CP}
\end{equation}
Then
\begin{equation} \label{GrindEQ__5_2_3_}
\begin{array}{l} {\, \, \left\| \frac{\tilde{\bPs }_{n} \left(\bth ^{*} \right)+\bPs _{n} \left(\bth ^{*} \right)-\tilde{\bPs }_{n} \left(\bth ^{*} \right)+\tilde{\bQ}\left(\hat{\bth }_{n} \right)}{n^{{-1 \mathord{\left/{\vphantom{-1 2}}\right.\kern-\nulldelimiterspace} 2} } +\left\| \tilde{\bQ}\left(\hat{\bth }_{n} \right)\right\| } \right\| } \\ {\le \tilde{\bZ}_{n} \left(\hat{\bth }_{n} ,\bth ^{*} \right)+\sqrt{n} \bPs _{n} \left(\hat{\bth }_{n} \right)+\sqrt{n} \left\{\left(\bPs _{n} \left(\hat{\bth }_{n} \right)-\tilde{\bPs }_{n} \left(\hat{\bth }_{n} \right)\right)-\left(\bPs _{n} \left(\bth ^{*} \right)-\tilde{\bPs }_{n} \left(\bth ^{*} \right)\right)\right\}} \\ {=\tilde{\bZ}_{n} \left(\hat{\bth }_{n} ,\bth ^{*} \right)+\sqrt{n} \bPs _{n} \left(\hat{\bth }_{n} \right)+\sqrt{n} \left(\bR\left(\hat{\bth }_{n} \right)-\bR\left(\bth ^{*} \right)\right)} \end{array}
\end{equation}
where $\bR\left(\bth \right)=\bPs _{n} \left(\bth \right)-\tilde{\bPs }_{n} \left(\bth \right)$ (this is parallel to (58) in Huber).

\noindent
Now, by the same arguments as in Lemma 3 of Huber, we have $\tilde{\bZ}_{n} \left(\hat{\bth }_{n} ,\bth ^{*} \right)
\mathop{\mathop{\to }\limits^{p} }\limits_{n\to \infty } 0$,
and by our Lemma 1, we have $\sqrt{n} \bPs _{n} \left(\hat{\bth }_{n} \right)
\mathop{\mathop{\to }\limits_{n\to \infty } }\limits^{p} 0$.

\noindent In regard to the third term :
\[\begin{array}{l} {\bR\left(\bth \right)=\frac{1}{n} \sum _{i=1}^{n}\left[\bvarphi _{i} \left(\bV_{i} ,\bth ,g\right)-\tilde{\bvarphi }_{i} \left(\bV_{i} ,\bth ,g\right)\right] } \\ {=\frac{1}{n} \sum _{i=1}^{n}\delta _{i} \left[\bxi _{i} \left(T_{i}^{0} ,\bth ,g\right)-\frac{S^{\left(1\right)} \left(T_{i}^{0} ,\bth ,g\right)}{S^{\left(0\right)} \left(T_{i}^{0} ,\bth ,g\right)} -\bxi _{i} \left(T_{i}^{0} ,\bth ,g\right)+\frac{s^{\left(1\right)} \left(T_{i}^{0} ,\bth ,g\right)}{s^{\left(0\right)} \left(T_{i}^{0} ,\bth ,g\right)} \right] } \\ {=-\frac{1}{n} \sum _{i=1}^{n}\delta _{i} \left[\frac{S^{\left(1\right)} \left(T_{i}^{0} ,\bth ,g\right)}{S^{\left(0\right)} \left(T_{i}^{0} ,\bth ,g\right)} -\frac{s^{\left(1\right)} \left(T_{i}^{0} ,\bth ,g\right)}{s^{\left(0\right)} \left(T_{i}^{0} ,\bth ,g\right)} \right]\,  .} \end{array}\]
We now need two intermediate results, which we call Lemma 4 and Lemma 5.

\noindent \textbf{Lemma 4}. We have
\begin{align*}
& \sqrt{n} \sup_{t, \bthsub} \left(S^{\left(j\right)} \left(t,{{\bth}} ,g\right)
- s^{\left(j\right)} \left(t,{{\bth}},g\right)\right) = O_p(1) \\
& \mathop{\sup }\limits_{t} \sqrt{n} \left[\left(S^{\left(j\right)} \left(t,{\hat{\bth}}_{n} ,g\right)-s^{\left(j\right)},
g\right)-s^{\left(j\right)} \left(t,{\hat{\bth}}_{n} ,g\right)\right)-\left(S^{\left(j\right)}
\left(t,\bth ^{*} ,g\right)-s^{\left(j\right)} \left(t,\bth ^{*} ,g\right)\right)]
\mathop{\mathop{\to }\limits_{n\to \infty } }\limits^{p} 0
\end{align*}
\noindent \textbf{Proof of Lemma 4. }We prove the Lemma for $j=0$, and the proof for $j=1$ is the same.
We have
\begin{align*}
S^{\left(0\right)} \left(t,\bth ,g\right)
& =\frac{1}{n} \sum _{i=1}^{n}Y_{i} \left(t\right) \exp \left(\bps ^{T} V_{i}
\left(g\left(t,\tau \right)\right)\right) \\
& =\frac{1}{n} \sum _{i=1}^{n}{\rm I}
\left(T_{i}^{0} \ge t\right) \exp \left(\bps ^{T} \bV_{i} \left(g\left(t,\tau \right)\right)\right)
= \mathbb{P}_{n} C_{\left(\bps ,\tau ,t\right)}^{}
\end{align*}
where $\mathbb{P}_{n}$ denotes the empirical measure and
$C_{\left(\bps ,\tau ,t\right)}^{} \left(s,v\right)={\rm I} \left(s\ge t\right)\exp \left(\bps ^{T} v\right)$.

\noindent Define $\Im =\left\{C_{\left(\bps ,\tau ,t\right)}^{} \, :\, \left(\bps ,\tau \right)\in \bTh ,\, t\in [0,t^{*}
]\right\}$.

\noindent (i) $\Im $ is a Donsker class:
Since $\bV\left(g\left(t\right)\right)$ is (approximately) bounded, and since $\bth $ is bounded by assumption GA1, then,
$\exp \left(\bth ^{T} v\right)$ is bounded. In addition, the indicator function ${\rm I} \left(s\ge t\right)$ is bounded (by 1).
Thus, $C_{\left(\bth ,t\right)}^{} \left(s,v\right)$ has the Lipschitz property,
and therefore it is Donsker by Example 19.7 in Van der Vaart (1998).
The first claim follows.

\noindent (ii) Define $d_{t}^{\left(n\right)} =C_{\left({\bhpsi},\hat{\tau },t\right)}^{} -C_{\left(\bps ^{*} ,\tau ^{*}
,t\right)}^{} $. Using the proof of Lemma 19.24 in Van der Vaart (1998) we have that $\sqrt{n}(\mathbb{P}_{n} - P)
d_{t}^{\left(n\right)} \to 0$ as a
process on $l^{\infty } \left([0,t^{*} ]\right)$. $\square$
\\
\noindent \textbf{Lemma 5.} We have
$$
\sqrt{n} \sup_t \left| \left[\left(\frac{S^{\left(1\right)} \left(t,\hat{\bth }_{n} ,g\right)}
{S^{\left(0\right)} \left(t,\hat{\bth }_{n} ,g\right)} -\frac{s^{\left(1\right)}
\left(t,\hat{\bth }_{n} ,g\right)}{s^{\left(0\right)} \left(t,\hat{\bth }_{n} ,g\right)} \right)
-\left(\frac{S^{\left(1\right)} \left(t,\bth ^{*} ,g\right)}{S^{\left(0\right)} \left(t,\bth ^{*} ,g\right)}
 -\frac{s^{\left(1\right)} \left(t,\bth ^{*} ,g\right)}
{s^{\left(0\right)} \left(t,\bth ^{*} ,g\right)} \right)\right] \right|
\mathop{\mathop{\to }\limits_{n\to \infty } }\limits^{p} 0
$$
\textbf{Proof of Lemma 5.}: This is a straightforward consequence of the preceding lemma. \\

\noindent We now continue with the proof of Theorem 1. Lemmas 4 and 5 yield $\sqrt{n} \left(\bR\left(\hat{\bth }_{n} \right)-\bR\left(\bth ^{*} \right)\right)=\left(-\frac{1}{n} \right)\sum _{i=1}^{n}\delta _{i} o_{p} \left(1\right) \le \left(-\frac{1}{n} \right)\sum _{i=1}^{n}o_{p} \left(1\right) =-o_{p} \left(1\right)\, \,.$

\noindent Thus all three terms on the right-hand side of (A.1) converge to 0 and so,
\\
$\, \left\| \frac{\tilde{\bPs }_{n} \left(\bth ^{*} \right)+[\bPs _{n} \left(\bth ^{*} \right)-\tilde{\bPs }_{n} \left(\bth ^{*} \right)]+\tilde{\bQ}\left(\hat{\bth }_{n} \right)}{n^{{-1 \mathord{\left/{\vphantom{-1 2}}\right.\kern-\nulldelimiterspace} 2} } +\left|\tilde{\bQ}\left(\hat{\bth }_{n} \right)\right|} \right\| \mathop{\mathop{\to }\limits^{p} }\limits_{n\to \infty } 0$.

\noindent Next, we can write
\[
\frac{S^{\left(1\right)} \left(t,\theta ^{*} ,g\right)}{S^{\left(0\right)} \left(t,\theta ^{*} ,g\right)}
-\frac{s^{\left(1\right)} \left(t,\theta ^{*} ,g\right)}{s^{\left(0\right)} \left(t,\theta ^{*} ,g\right)}
=\frac{1}{S^{\left(0\right)} \left(t,\theta ^{*} ,g\right)} \left[S^{\left(1\right)} \left(t,\theta ^{*}
,g\right)-\frac{s^{\left(1\right)} \left(t,\theta ^{*} ,g\right)}{s^{\left(0\right)} \left(t,\theta ^{*} ,g\right)}
S^{\left(0\right)} \left(t,\theta ^{*} ,g\right)\right].
\]
By Lemma 4, the last expression is asymptotically equivalent to
\begin{align*}
\frac{1}{s^{\left(0\right)} \left(t,\theta ^{*}
,g\right)} & \left[S^{\left(1\right)} \left(t,\theta ^{*} ,g\right)-\frac{s^{\left(1\right)} \left(t,\theta ^{*}
,g\right)}{s^{\left(0\right)} \left(t,\theta ^{*} ,g\right)} S^{\left(0\right)} \left(t,\theta ^{*} ,g\right)\right] \\
& = \frac{1}{s^{\left(0\right)} \left(t,\theta ^{*} ,g\right)} \left[\frac{1}{n} \sum _{i=1}^{n}Y_{i}
\left(T_{i}^{0} \right) \xi _{i}
\left(T_{i}^{0} ,\theta ^{*} ,g\right)\exp \left(\psi ^{T} V_{i} \left(T_{i}^{0} ,\theta ^{*} ,g\right)\right) \right. \\
& \hspace*{36pt} \left. -\frac{s^{\left(1\right)} \left(t,\theta ^{*} ,g\right)}{s^{\left(0\right)} \left(t,\theta ^{*} ,g\right)} \frac{1}{n}
\sum_{i=1}^{n}Y_{i} \left(t\right)\exp \left(\psi ^{T} V_{i} \left(T_{i}^{0} ,\theta ^{*} ,g\right)\right) \right]
\end{align*}
Therefore,
\begin{equation}
\tilde{\bPs}_{n} \left(\bth ^{*} \right)-\bPs_{n} \left(\bth ^{*} \right)
=\frac{1}{n} \sum _{i=1}^{n}\tilde{h}_{i}  \left(\theta ^{*} \right)
+ o_{p} \left(n^{-{1 \mathord{\left/{\vphantom{12}}\right.\kern-\nulldelimiterspace} 2} } \right)
\label{psi}
\end{equation}
where
\[\begin{array}{l} {\tilde{h}_{i} \left(\theta ^{*} \right)\equiv \tilde{h}\left(V_{i} ,\theta ^{*} ,g\right)} \\
{=\frac{\delta _{i} }{s^{\left(0\right)} \left(T_{i}^{0} ,\theta ^{*} ,g\right)} \left[\begin{array}{l} {\frac{1}{n} \sum
_{i=1}^{n}Y_{i} \left(T_{i}^{0} \right)\xi _{i} \left(T_{i}^{0} ,\theta ^{*} ,g\right)\exp \left(\psi ^{T} V_{i} \left(T_{i}^{0}
,\theta ^{*} ,g\right)\right) } \\ {-\frac{s^{\left(1\right)} \left(T_{i}^{0} ,\theta ^{*} ,g\right)}{s^{\left(0\right)}
\left(T_{i}^{0} ,\theta ^{*} ,g\right)} \frac{1}{n} \sum _{i=1}^{n}Y_{i} \left(T_{i}^{0} \right)\exp \left(\psi ^{T} V_{i}
\left(T_{i}^{0} ,\theta ^{*} ,g\right)\right) } \end{array}\right]\, }
\end{array}\]

\noindent Thus,
\[\begin{array}{l} {\, \left\| \frac{\tilde{\bPs }_{n} \left(\bth ^{*} \right)+\bPs _{n} \left(\bth ^{*} \right)-\tilde{\bPs }_{n} \left(\bth ^{*} \right)+\tilde{\bQ}\left(\hat{\bth }_{n} \right)}{n^{{-1 \mathord{\left/{\vphantom{-1 2}}\right.\kern-\nulldelimiterspace} 2} } +\left\| \tilde{\bQ}\left(\hat{\bth }_{n} \right)\right\| } \right\| =\left\| \frac{\tilde{\bPs }_{n} \left(\bth ^{*} \right)+\frac{1}{n} \sum _{i=1}^{n}\tilde{h}_{i} \left(\bth^{*} \right)+ \tilde{\bQ}\left(\hat{\bth }_{n} \right)}{n^{{-1 \mathord{\left/{\vphantom{-1 2}}\right.\kern-\nulldelimiterspace} 2} } +\left\| \tilde{\bQ}\left(\hat{\bth }_{n} \right)\right\| } \right\| } \\ {=\left\| \frac{\frac{1}{n} \sum _{i=1}^{n}\left(\tilde{\bvarphi }_{i} \left(\bth ^{*} \right)+\tilde{h}_{i} \left(\bth ^{*} \right)\right)+ \tilde{\bQ}\left(\hat{\bth }_{n} \right)}{n^{{-1 \mathord{\left/{\vphantom{-1 2}}\right.\kern-\nulldelimiterspace} 2} } +\left\| \tilde{\bQ}\left(\hat{\bth }_{n} \right)\right\| } \right\| =\left\| \frac{\frac{1}{n} \sum _{i=1}^{n}\tilde{\tilde{\bvarphi }}_{i} \left(\bth ^{*} \right)+ \tilde{\bQ}\left(\hat{\bth }_{n} \right)}{n^{{-1 \mathord{\left/{\vphantom{-1 2}}\right.\kern-\nulldelimiterspace} 2} } +\left\| \tilde{\bQ}\left(\hat{\bth }_{n} \right)\right\| } \right\| } \end{array}\]
and since the left-hand side converges to 0, so does the right-hand side (parallel to Eqn.\ (58) in Huber (1967)).
From this point, using the same arguments as in Huber (1967), we can obtain equations parallel to
Huber's Eqns.\ (59)-(62) and the desired result follows. $\square$ \\

\noindent \textbf{Proof of} \textbf{Corollary 1. }We can write
\[-\sqrt{n} \cdot \tilde{\bQ}\left({\hat{\bth}}_{n} \right)=-\left(\frac{1}{\sqrt{n} } \sum _{i=1}^{n}\tilde{\tilde{\bvarphi}}_{i} \left(\bth ^{*} \right)+\sqrt{n} \cdot \tilde{\bQ}\left({\hat{\bth}}_{n} \right) \right)+\frac{1}{\sqrt{n} } \sum _{i=1}^{n}\tilde{\tilde{\bvarphi}}_{i} \left(\bth ^{*} \right) \, .\]
By Theorem 1, the first term tends to zero in probability as $n\to \infty $. Since $\sum _{i=1}^{n}\tilde{\tilde{\bvarphi}}_{i}
\left(\bth ^{*} \right) $ is a sum of i.i.d. terms, we can apply the central limit theorem. We noted previously (in the remark
after Theorem 1 in the main text) that $E\left(\tilde{\tilde{\bps }}\left(\bth ^{*} \right)\right)=0$. Therefore,
$\frac{1}{\sqrt{n} } \sum _{i=1}^{n}\tilde{\tilde{\bvarphi}}_{i} \left(\bth ^{*} \right)$ $\mathop{\to }\limits^{d}
N\left(0,\bC\left(\bth ^{*} \right)\right)$, and so $\sqrt{n} \cdot \tilde{\bQ}\left({\hat{\bth}}_{n} \right)\mathop{\to
}\limits^{d} N\left(0,\bC\left(\bth ^{*} \right)\right)$.
Next, using the fact that $\tilde{\bQ}\left(\bth ^{*} \right)=0$ and carrying out a Taylor expansion, we have
\[
\tilde{\bQ}\left({\hat{\bth}}\right) = \tilde{\bQ}\left({\hat{\bth}}\right)-\tilde{\bQ}\left(\bth ^{*} \right)=\bLam \left({\hat{\bth}}-\bth ^{*} \right)
+o\left(\left\| {\hat{\bth}}-\bth ^{*} \right\| \right).
\]
Multiplying both sides by $\sqrt{n} \bLam ^{-1} $ yields:
$$
\sqrt{n} \bLam ^{-1} \tilde{\bQ}\left({\hat{\bth}}_{n} \right)
=\sqrt{n} \left({\hat{\bth}}_{n} -\bth ^{*} \right)+\sqrt{n} c_{n} \left({\hat{\bth}}_{n} -\bth ^{*} \right)
=\sqrt{n} \left({\hat{\bth}}_{n} -\bth ^{*} \right)\left(1+c_{n} \right),
$$
where $c_{n} \mathop{\to }\limits^{p} 0$. Thus $\sqrt{n}
\left({\hat{\bth}}_{n} -\bth ^{*} \right)\mathop{\to }\limits^{d} \bLam ^{-1} \cdot N\left(0,\bC\left(\bth ^{*}
\right)\right)=N\left(0,\bLam ^{-1} \bC\left(\bth ^{*} \right)\left(\bLam ^{-1} \right)^{T} \right)$. $\square$

\newpage
\clearpage


\end{document}